\pdfoutput=1
\documentclass[aps,prl,superscriptaddress,
twocolumn,
]{revtex4-2}
\usepackage[dvipdfmx]{graphicx}
\usepackage[svgnames,psnames]{xcolor}
\usepackage[T1]{fontenc}
\usepackage{textcomp}
\usepackage{bm}
\usepackage{amsmath,amssymb,amsthm}
\usepackage{mathtools}
\usepackage{mathrsfs}
\usepackage{comment}
\usepackage[normalem]{ulem}
\usepackage{braket}
\usepackage{siunitx}

\newcommand{\tr}{\mathrm{tr}\,}

\newcommand{\Abs}[1]{\left|#1\right|}

\usepackage[
colorlinks=true,
citecolor=DarkGreen,
linkcolor=FireBrick,
unicode
]{hyperref}

\allowdisplaybreaks[1]
\begin{document}
%

\title{Disorder-Induced Slow Relaxation of Phonon Polarization}

\author{Yuta Suzuki}
\email{suzuki.y.8cc2@m.isct.ac.jp}
\altaffiliation[Previous address: ]{Department of Physics, The University of Tokyo, 7-3-1 Hongo, Bunkyo-ku, Tokyo 113-0033, Japan}
\affiliation{Department of Physics, Institute of Science Tokyo, 2-12-1 Ookayama, Meguro-ku, Tokyo 152-8551, Japan}
\author{Shuichi Murakami}
\altaffiliation[Previous address: ]{Department of Physics, Tokyo Institute of Technology, 2-12-1 Ookayama, Meguro-ku, Tokyo 152-8551, Japan}
\affiliation{Department of Physics, Institute of Science Tokyo, 2-12-1 Ookayama, Meguro-ku, Tokyo 152-8551, Japan}
\affiliation{International Institute for Sustainability with Knotted Chiral Meta Matter (WPI-SKCM${}^\text{2}$),
Hiroshima University, 1-3-1 Kagamiyama, Higashi-Hiroshima, Hiroshima 739-8526, Japan}


\date{\today}

\begin{abstract}%
The role of the polarization degree of freedom in lattice dynamics in solids has been underlined recently. 
We theoretically discover a relaxation mechanism for both linear and circular polarizations of acoustic phonons. 
In the absence of scattering, the polarization exhibits oscillatory behavior. 
This behavior leads to a counterintuitive result: unlike linear momentum, more frequent scattering events cause slower polarization relaxation due to motional narrowing. 
We validate this mechanism using the quantum kinetic equation.  
We derive the relaxation rates of polarizations analytically for isotropic elastic bodies and numerically for a cubic crystal.
Remarkably, we reveal that linear polarizations relax more slowly than circular ones. 
Our findings provide a pathway to extend the lifetime of phonon angular momentum.
\end{abstract}
\maketitle

\textit{\textcolor[named]{DarkGreen}{Introduction}}.---The discovery that phonons, quantized lattice vibrations, can carry angular momentum 
(AM)~\cite{Vonsovskii1962,Levine1962,Portigal1968,Pine1970,Ishii1975,Mclellan1988,Zhang2014,Zhang2015,Chen2015,Zhu2018,Kishine2020,
Zhang2022,Chen2015,Zhu2018,AKato2023,Ishito2023a,Ishito2023b,Oishi2024}
besides heat and linear momentum has enriched our understanding of 
transport in crystals~\cite{Hamada2018,Hamada2020a,Park2020,Zhang2021,Chen2022,Yao2022,Romao2023}. 
In particular, the phonon AM, arising from the imbalance between right- and left-circularly polarized phonons, 
can couple to electron spins and magnons~\cite{Korenev2016,Nova2017,Holanda2018,Sasaki2021,Jeong2022,Kim2023,Ueda2023,Ohe2024,
Juraschek2019,Hamada2020,Juraschek2022,Tauchert2022,Fransson2023,Chaudhary2024,Funato2024,Yao2024}.

The phonon AM can be generated by thermal gradients~\cite{Hamada2018} 
or interfacial injection~\cite{Korenev2016,Nova2017,Holanda2018,Sasaki2021,Jeong2022,Kim2023,Ohe2024,Suzuki2024}
and vanishes in equilibrium in nonmagnetic crystals~\cite{Zhang2014,Lin2024}.
This indicates that the AM is lost during thermalization.
Meanwhile, the relaxation process of the phonon AM remains unexplored. 
The key questions are what determines the relaxation timescale and how it depends on disorder strength. 
Resolving these issues would facilitate experimental observations and applications of chiral phonons.

In this Letter, we theoretically propose a relaxation mechanism for the phonon AM via polarization oscillations between collisions and resulting phase relaxation.
This mechanism, analogous to the Dyakonov--Perel mechanism~\cite{Gershenzon1969,Dyakonov1971a,Dyakonov1972,Fabian2007}
for electron spins, arises from scattering by isotropic defects, 
such as atoms with different masses or point defects~\cite{Klemens1955,Tamura1983,Srivastava1990,ZimanTextbookElPh}.
Remarkably, the AM relaxation time $\tau^{\text{AM}}$ becomes inversely proportional to the linear-momentum relaxation time $\tau^*$
($\tau^{\text{AM}}\propto 1/\tau^*$).
Namely, this mechanism predicts that stronger disorder enhances robustness of AM.
This contrasts with a na\"ive expectation that 
an AM relaxation time would be proportional to the linear-momentum relaxation time ($\tau^{\text{AM}}\propto \tau^*$)~\footnote{
For electron spins, a similar process where spin-flip scattering from impurities accelerates AM relaxation
is known as the Elliott--Yafet mechanism~\cite{Elliott1954,Yafet1963,Fabian2007}. 
}.

\begin{figure}
 \centering
\includegraphics[pagebox=artbox,width=0.90\columnwidth]{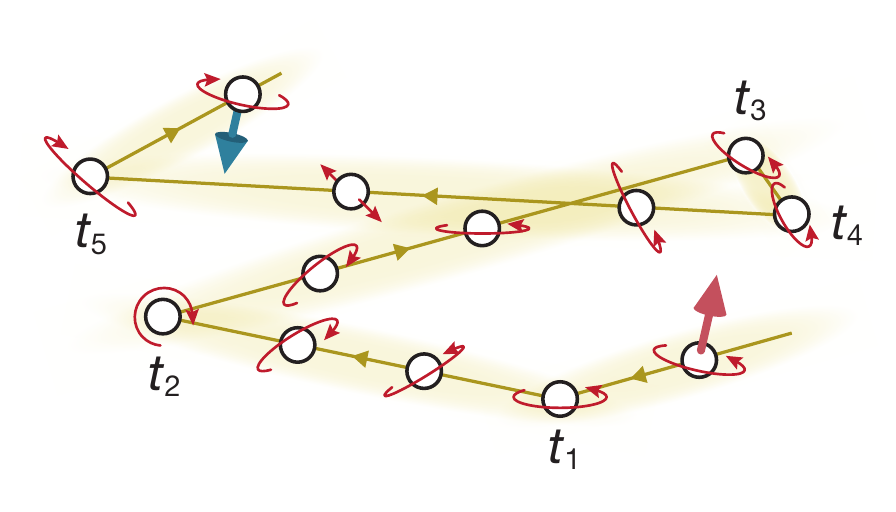}
\caption{
Schematic of a relaxation mechanism for phonon polarization. 
Thin red arrows on white spheres represent the polarizations, while 
thick arrows denote phonon angular momenta (AM), which are reversed between the initial and final states.
Phonons acquire random linear momenta due to scattering at times $t = t_1, t_2, \dots$, following the yellow zigzag path.
As a result, the polarization experiences random kicks, which lead to a slow relaxation analogous to motional narrowing.}
\label{fig: schematics mational narrowing}
\end{figure}

We provide a schematic of the relaxation mechanism in Fig.~\ref{fig: schematics mational narrowing}.
Let us focus on circularly-polarized acoustic phonons, which are thermally accessible. 
In crystals, there exists a band splitting between longitudinal and transverse modes.
It implies the presence of a Hamiltonian term, ${H}^{\prime}_{\bm{q}}$ [Eq.~\eqref{eq: general H mat}], that lifts the degeneracy of polarization states. 
This term acts as a wavevector-dependent effective field for phonon polarizations,
analogous to spin--orbit coupling for electron spins, and causes the circular polarization~(CP) to oscillate. 
Upon scattering at times $t = t_1, t_2, \dots$, the wavevector $\bm{q}$ and consequently ${H}^{\prime}_{\bm{q}}$ change,
introducing random fluctuations to the oscillation.
When the scattering rate is high, such that the interval between collisions is shorter than the oscillation period, 
the fluctuations are averaged out in the long-term behavior.
Thus, crystal disorder impedes the relaxation of CP.
This scenario is analogous to the motional narrowing of the linewidth in NMR~\cite{Bloembergen1948,Anderson1953,Vleck1948,Kubo1954,Kubo1991,Kittel2004}.

Our framework applies not only to CPs but to general polarization states of phonons. 
With analytical calculations for isotropic elastic bodies and numerical simulations for a cubic crystal,
we reveal that linear polarizations~(LPs) tend to relax more slowly than CPs,
though all polarizations have relaxation times of the same order. 

Hereafter, we focus on acoustic modes of nonmagnetic centrosymmetric crystals to illustrate this mechanism.

\textit{\textcolor[named]{DarkGreen}{Polarization parameters}}.---We first characterize the LPs and CPs of acoustic phonons.
Let us consider a plane wave of phonons in the elastic limit, with displacement $2\mathrm{Re}[\bm{u} \exp (i \bm{q}\cdot \bm{r} - i\omega t)]$,
where $\bm{u}$, $\bm{q}$, and $\omega$ are the amplitude, wavevector, and frequency, respectively.
By using the polarization vector $\bm{e} \equiv \bm{u}/ \Abs{\bm{u}} = (e_x, e_y, e_z)$, 
AM associated with the displacement is given by $\bm{e}^{\dagger} {S}_{\mu}\bm{e}$~\cite{Vonsovskii1962,Levine1962,Ishii1975,Mclellan1988,Zhang2014},
where ${\bm{S}} = ({S}_x, {S}_y, {S}_z)$ are $3\times 3$ Hermitian matrices of spin-1 operators 
with elements $({S}_{\mu})_{\nu, \kappa}= -i\hbar \epsilon_{\mu\nu\kappa}$ ($\mu, \nu, \kappa = x, y, z$)~\footnote{
If we define phonon AM by the statistical average of mechanical AM of all atoms~\cite{Vonsovskii1962,Levine1962,Ishii1975,Mclellan1988,Zhang2014}, 
other interband term appears off equilibrium~\cite{Zhang2014,Zhong2023}.
However, in isotropic elastic bodies, the AM relaxation time $\tau^{\text{AM}}$ remains unaffected by this correction term~\cite{SupplementalMaterial}.
}.

\begin{figure}
 \centering
\includegraphics[pagebox=artbox,width=0.99\columnwidth]{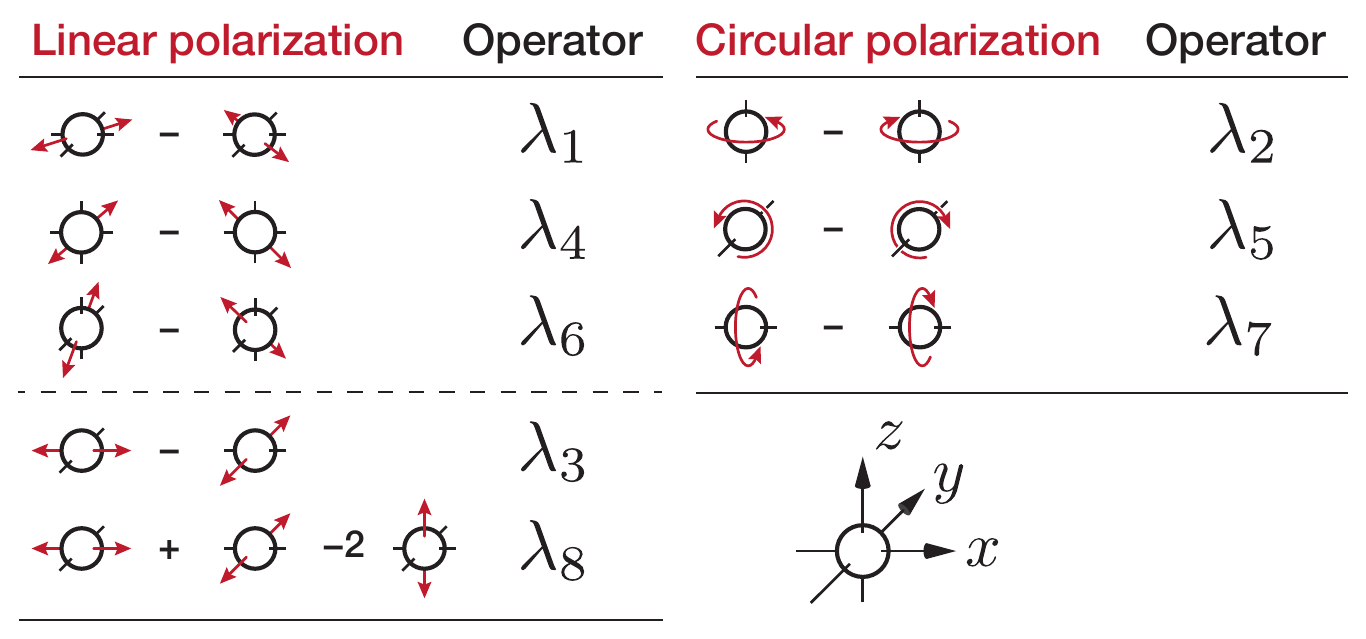}
\caption{
Schematic list of eight parameters $\bm{e}^{\dagger}{\lambda}_a \bm{e}$,
which represent expectation values of the Gell-Mann matrices ${\lambda}_a$ with polarization vector $\bm{e}$.
These parameters fully characterize polarization states of long-wavelength acoustic phonons.
Thin red arrows on white spheres denote linear polarization~(LP) or circular polarization~(CP). 
}
\label{fig: phon pol table and motional narr}
\end{figure}

In addition to the CP characterized by $\bm{e}^{\dagger} {S}_{\mu}\bm{e}$, we introduce five new components for phonon polarization.
Since any $3\times 3$ Hermitian matrix can be expanded in terms of the identity matrix ${I}_3$ and 
the Gell-Mann matrices~\cite{Gell-Mann1962}, denoted by ${\lambda}_a$ $(a = 1, 2,\dots, 8)$,
we introduce eight parameters $\bm{e}^{\dagger}{\lambda}_a \bm{e}$.
These parameters have been used as generalized Stokes parameters in optics~\cite{Roman1959,Barakat1977,Setala2002a} and geophysics~\cite{Samson1973},
but received little attention in crystal acoustics, except for Ref.~\cite{Sugic2020}.
In Fig.~\ref{fig: phon pol table and motional narr}, we illustrate the degrees of polarization corresponding to $\bm{e}^{\dagger}{\lambda}_a \bm{e}$.
The five of eight parameters, characterized by symmetric matrices ${\lambda}_a$ with indices $a \in \{1,3,4,6,8\}\equiv \mathsf{S}$,
measure $d$-orbital-like linear polarizations, while 
the other three parameters, characterized by antisymmetric matrices ${\lambda}_a$ with indices $a \in \{2,5,7\}\equiv \mathsf{AS}$,
measure phonon AM.
Among the five parameters with
$a\in \mathsf{S}$, three parameters define the degree of LP inclined at $\pi/4$ angle
in the $xy$-plane ($a =1$), $zx$-plane ($a =4$), and $yz$-plane ($a =6$).
The other two parameters describe the polarization difference between the $x$- and $y$-directions ($a =3$)
and the difference between in-plane and out-of-plane polarization with respect to the $xy$-plane ($a =8$).
The three parameters with $a\in \mathsf{AS}$ correspond to the spin-1 operators: 
${S}_{x} = \hbar{\lambda}_7$, ${S}_{y}= -\hbar{\lambda}_5$, and ${S}_z= \hbar{\lambda}_2$.

\textit{\textcolor[named]{DarkGreen}{Undamped dynamics}}.---The equation of motion for the displacement results in an eigenvalue problem
${D}(\bm{q}) \bm{u}  = \omega^2\bm{u}$, where ${D}(\bm{q})$ is the dynamical matrix.
We introduce the Hamiltonian ${H}_{\bm{q}}$ for acoustic modes as a square root of ${D}(\bm{q})$.
In nonmagnetic centrosymmetric crystals, ${H}_{\bm{q}}$ is expanded in terms of symmetric matrices~\cite{SupplementalMaterial}:
\begin{equation}
{H}_{\bm{q}} = h_0(\bm{q})\cdot{I}_3 + {H}^{\prime}_{\bm{q}},\qquad
{H}^{\prime}_{\bm{q}}=\sum_{a \in \mathsf{S}} h_a (\bm{q}) {\lambda}_a.
\label{eq: general H mat}
\end{equation}
Here, $h_0(\bm{q})$ and $h_{a} (\bm{q})$ are real functions of $\bm{q}$.

Let us consider dynamics of the polarizations under ${H}_{\bm{q}}$.
We describe mixed states of phonons by the $3\times 3$ density matrix ${\varrho}_{\bm{q}}$~\cite{PeierlsTextbook,Simoncelli2019,Zhong2023}.
The density matrix becomes ${\varrho}^{\text{eq}}_{\bm{q}}= 
\left[\exp ({H}_{\bm{q}}/k_{B}T) - I_3\right]^{-1}$ in equilibrium at temperature $T$ and
deviates from it off equilibrium. 
We express the deviation as ${\rho}_{\bm{q}}\equiv {\varrho}_{\bm{q}} - {\varrho}^{\text{eq}}_{\bm{q}}$.
With no collisions, its dynamics is governed by the von~Neumann equation
$\partial {\rho}_{\bm{q}}/\partial t = - ({i}/{\hbar})\left[{H}^{\prime}_{\bm{q}},{\rho}_{\bm{q}}\right]$.
Each polarization~\cite{Setala2002a,Sugic2020}
$\Lambda^a_{\bm{q}} \equiv (3/2)\tr [{\rho}_{\bm{q}} {\lambda}_{a}]$ at a fixed value of $\bm{q}$ thus follows 
\begin{subequations}
 \begin{align}
 \frac{d \Lambda^a_{\bm{q}}}{dt} &= \frac{2}{\hbar} \sum_{c\in \mathsf{AS}} \sum_{b \in \mathsf{S}} f_{abc} h_b(\bm{q}) \Lambda^c_{\bm{q}}  &a& \in \mathsf{S},
 \label{eq: coupled osc S and  AS 1} \\
 \frac{d \Lambda^c_{\bm{q}}}{dt} &= -\frac{2}{\hbar} \sum_{a\in \mathsf{S}} \sum_{b \in \mathsf{S}} f_{abc} h_b(\bm{q}) \Lambda^a_{\bm{q}}  &c& \in \mathsf{AS}.
 \label{eq: coupled osc S and  AS 2}
\end{align}
\end{subequations}
Here, we used a relation $\left[{\lambda}_a, {\lambda}_b\right] = 2i \sum_{c = 1}^8 f_{abc} {\lambda}_c$
and properties of the structure constant $f_{abc}$~\cite{SupplementalMaterial}.

Equations~\eqref{eq: coupled osc S and AS 1}\eqref{eq: coupled osc S and AS 2}
describe coupled dynamics between linear and rotational deformations, and yield an oscillatory solution for the polarizations.
A similar analysis was conducted for electron $p$-orbital dynamics~\cite{Han2022}.
We can also interpret this oscillation as a motion of spherical pendulum in the eight-dimensional space,
since the point $(\Lambda^1_{\bm{q}},\Lambda^2_{\bm{q}}, \dots, \Lambda^8_{\bm{q}})$ 
traces a hypersphere of radius 
$\sqrt{\sum_{a = 1}^8 (\Lambda^a_{\bm{q}})^2}$~\cite{SupplementalMaterial}.
This is an analog of precession of an electron spin in the three-dimensional space.

\textit{\textcolor[named]{DarkGreen}{Kinetic theory}}.---We then consider 
relaxation process of the polarization under frequent scattering events.
We use the quantum kinetic equation~\cite{Kohn1957,Dyakonov1971a,Dyakonov1972,Ivchenko1989,Ivchenko1990,Shytov2006,Kailasvuori2009,Sohn2024}
$ \partial {\rho}_{\bm{q}}/{\partial t} + ({i}/{\hbar})\left[{H}^{\prime}_{\bm{q}}, {\rho}_{\bm{q}}\right] 
= \mathrm{St}[{\rho}]$
with the collision integral due to random elastic scattering by isotropic scatterers~\footnote{
To be exact, the collision term consists of symmetrized product of the matrix delta function $\delta(H_{\bm{q}} - H_{\bm{q}'})$ and the density matrix 
difference $(\rho_{\bm{q}} - \rho_{\bm{q}'})$~\cite{Ivchenko1990}.
While our approximation in Eq.~\eqref{183542_15Aug24} is not universally valid---since 
degeneracy between longitudinal and transverse branches is lifted---it 
is widely used for the analysis of motional narrowing in electron spin~\cite{Dyakonov1971a,Dyakonov1972} and orbital AM~\cite{Sohn2024}.
We expect that Eq.~\eqref{183542_15Aug24} yields qualitatively consistent relaxation times.}
\begin{equation}
 \mathrm{St}[{\rho}] = \frac{2\pi}{\hbar} \sum_{\bm{q}'}\mathcal{K}_{\bm{q},\bm{q}'}
\left({\rho}_{\bm{q}'} - {\rho}_{\bm{q}}\right) \cdot \delta (\hbar\overline{\omega}_{q} - \hbar\overline{\omega}_{q'}).
\label{183542_15Aug24} 
\end{equation}
Here, $q = \Abs{\bm{q}}$, $q' = \Abs{\bm{q}'}$, and $\mathcal{K}_{\bm{q},\bm{q}'}$ denotes the scattering rate.
We defined a dispersion of an isotropic effective branch $\hbar\overline{\omega}_{q}$ by
averaging $h_0(\bm{q})$ over all directions $\hat{\bm{q}} = \bm{q}/q$.
In Eq.~\eqref{183542_15Aug24},
we assume both the isotropy of the scattering potential and its independence from polarization degrees of freedom.
In other words, we consider that $\mathcal{K}_{\bm{q},\bm{q}'}$ is a scalar quantity, independent of polarization, and 
depends only on the energy $\hbar\overline{\omega}_{q}$ and the angle between $\hat{\bm{q}}$ and $\hat{\bm{q}}'$.

Let us rewrite the quantum kinetic equation for the case of isotropic elastic bodies.
The Hamiltonian~\eqref{eq: general H mat} of the isotropic body 
with longitudinal~($c_{\text{L}}$) and transverse~($c_{\text{T}}$) sound velocities is expressed as
$ \left({H}_{\bm{q}}\right)_{\mu\nu}
= \hbar c_{\text{T}} q \left( \delta_{\mu\nu} - \hat{q}_{\mu}\hat{q}_{\nu}\right) + \hbar c_{\text{L}} q \cdot \hat{q}_{\mu}\hat{q}_{\nu}$,
where $\mu, \nu= x, y, z$. 
Within the Hamiltonian, the scalar part is given as
$h_0(\bm{q}) = \hbar \overline{\omega}_q = \hbar v q$ with a mean velocity $v = (c_{\text{L}} + 2c_{\text{T}})/3$,
while the traceless part is given as
\begin{equation}
\left({H}^{\prime}_{\bm{q}}\right)_{\mu\nu}
= \hbar (c_{\text{L}} - c_{\text{T}}) q\left(\hat{q}_{\mu}\hat{q}_{\nu} - \frac{\delta_{\mu\nu}}{3}\right). 
\end{equation}
We also expand the density matrix ${\rho}_{\bm{q}}(t)$ using the spherical harmonics $Y_{lm}(\hat{\bm{q}})$ and the Gell-Mann matrices ${\lambda}_a$ as 
${\rho}_{\bm{q}}(t) = {\rho}^0_{\bm{q}}(t) \cdot {I}_{3} + \sum_{a=1}^8 \sum_{l=0}^\infty \sum_{m=-l}^l {\rho}^{a}_{lm}(q, t) Y_{lm}(\hat{\bm{q}}) {\lambda}_a$.
Here, ${\rho}^{a}_{lm}(q, t)$ represents the multipole components.
Then, it satisfies the following equation~\cite{SupplementalMaterial}:
\begin{equation}
\frac{\partial {\rho}^{a}_{lm}}{\partial t}
= \sum_{b, l', m'} i\Omega^{a;b}_{lm;l'm'} {\rho}^{b}_{l'm'}  - \frac{{\rho}^a_{lm}}{\tau_l}.
\label{094314_26Nov24}
\end{equation}
On the right-hand side, the first and second terms correspond to the commutator and collision integral of the quantum kinetic equation, respectively. 
The coefficients $\Omega^{a;b}_{lm;l'm'}$ are proportional to 
the band splitting $(c_{\text{L}}-c_{\text{T}}) q$, and they couple the multipole components between $a\in \mathsf{S}$ and $b\in \mathsf{AS}$ or vice versa. 
This coupling induces polarization oscillations, as shown in Eqs.~\eqref{eq: coupled osc S and AS 1}\eqref{eq: coupled osc S and AS 2}.
In contrast, the relaxation times $\tau_l$ for the $l$-th spherical harmonics of ${\rho}_{\bm{q}}$
introduce dissipation to the phonon system.

We now focus on the dynamics of $\hat{\bm{q}}$-independent components ${\rho}^a_{00}$~\footnote{
Similar discussion for electron spins in terms of the Dyakonov--Perel mechanism~\cite{Gershenzon1969,Dyakonov1971a,Dyakonov1972} 
is found in Refs.~\cite{Ivchenko1990,Fabian2007}.
}, since they yield a statistical average of the polarization, defined by
\begin{equation}
\sum_{\bm{q}} \Lambda^a_{\bm{q}} = \frac{3}{2\sqrt{\pi}}\int^{\infty}_0 d\overline{\omega}_q~g (\overline{\omega}_q)\cdot 
\rho^a_{00}(q, t).
\label{eq: phonon polarization density}  
\end{equation}
Here, $g(\overline{\omega}_q)$ is the density of states. 
Let us introduce two vectors $\bm{\rho}^{\mathsf{AS}}_{00}(q, t) = (\rho^2_{00}, \rho^5_{00}, \rho^7_{00})^{\top}$ 
and $\bm{\rho}^{\mathsf{S}}_{00}(q, t) = (\rho^1_{00}, \rho^3_{00}, \rho^4_{00}, \rho^6_{00}, \rho^8_{00})^{\top}$.
The former yields the expectation values of CPs, while the latter yields those of LPs.
According to Eq.~\eqref{094314_26Nov24}, their dynamics are rewritten as
\begin{equation}
 \frac{\partial}{\partial t} 
 \left[
\renewcommand\arraystretch{1.3}
\begin{array}{c}
 \bm{\rho}^{\mathsf{AS}}_{00} \\
 \bm{\rho}^{\mathsf{S}}_{00} \\ \hline
 \bm{\rho}^{\mathsf{AS}}_{2} \\
 \bm{\rho}^{\mathsf{S}}_{2} \\
\end{array} 
\right]  \simeq 
 \left[
\renewcommand\arraystretch{1.3}
\begin{array}{cc|cc}
  O                 & O                              & O                     & i{\Omega}           \\
  O                 & O                              & i\widetilde{{\Omega}}   & O                \\ \hline
  O                 & i\widetilde{{\Omega}}^{\dagger}& -\dfrac{1}{\tau_2}     & i{\Omega}'           \\
  i{\Omega}^{\dagger} & O                              & i{{\Omega}'}^{\dagger}  & -\dfrac{1}{\tau_2}\\
\end{array} 
\right] \left[
\renewcommand\arraystretch{1.3}
\begin{array}{c}
 \bm{\rho}^{\mathsf{AS}}_{00} \\
 \bm{\rho}^{\mathsf{S}}_{00} \\ \hline
 \bm{\rho}^{\mathsf{AS}}_{2} \\
 \bm{\rho}^{\mathsf{S}}_{2} \\
\end{array} 
\right].
 \label{112102_16Aug24}
\end{equation}
Here, $\bm{\rho}^{\mathsf{AS}}_{2}$ (or $\bm{\rho}^{\mathsf{S}}_{2}$) denotes a vector of multipole components $\rho^{a}_{2 m}(q, t)$ 
with $a \in \mathsf{AS}$ (or $a \in \mathsf{S}$) coupled to the dynamics of $\rho^a_{00}(q, t)$.
We omitted other multipole components $\rho^a_{lm}$ with $l\neq 0, 2$ appearing in Eq.~\eqref{112102_16Aug24},
since these components do not directly couple to the dynamics of $\rho^a_{00}$
whereas they may couple to that of $\bm{\rho}^{\mathsf{AS}/\mathsf{S}}_{2}$.
The submatrices ${\Omega}$, $\widetilde{{\Omega}}$, and ${{\Omega}'}$ 
consist of the coefficients $\Omega^{a;b}_{lm;l'm'}$~\cite{SupplementalMaterial}.

Let us consider a region where the scattering rate $1/\tau_2$ exceeds the typical oscillation frequency 
$\Omega^{a;b}_{lm;l'm'} \simeq (c_{\text{L}}- c_{\text{T}})q$ at a fixed value of $q$.
In this region, the components $\rho^a_{00}$ relax slowly: 
they remain invariant under polarization-conserving scatterings and experience the perturbations during brief intervals between scattering events.
In contrast, the multipole components $\bm{\rho}^{\mathsf{AS/S}}_{2}$ relax quickly to quasistatic distribution at a rate $1/\tau_2$.
We obtain the quasistatic distribution by setting 
$ \partial \bm{\rho}^{\mathsf{AS}}_{2}/\partial t = \partial \bm{\rho}^{\mathsf{S}}_{2}/\partial t = 0 $
on the left-hand side of Eq.~\eqref{112102_16Aug24}.
This yields
$\bm{\rho}^{\mathsf{AS}}_{2} \simeq i\widetilde{{\Omega}}^{\dagger} \tau_2 \bm{\rho}^{\mathsf{S}}_{00}$ and 
$\bm{\rho}^{\mathsf{S}}_{2} \simeq i{{\Omega}}^{\dagger} \tau_2 \bm{\rho}^{\mathsf{AS}}_{00}$.

We finally determine the slow dynamics of $\rho^a_{00}$ by inserting the quasistatic distribution into Eq.~\eqref{112102_16Aug24}:
\begin{subequations}
 \begin{align}
 \frac{\partial \bm{\rho}^{\mathsf{AS}}_{00}}{\partial t} 
 &\simeq - {\Omega}{\Omega}^{\dagger}\tau_2\cdot \bm{\rho}^{\mathsf{AS}}_{00} = - \frac{\bm{\rho}^{\mathsf{AS}}_{00}}{\tau^{\text{AM}}}, \label{195042_13Dec24}\\
 \frac{\partial\bm{\rho}^{\mathsf{S}}_{00}}{\partial t} 
 &\simeq - \widetilde{{\Omega}}\widetilde{{\Omega}}^{\dagger} \tau_2\cdot \bm{\rho}^{\mathsf{S}}_{00}
 = - \frac{\bm{\rho}^{\mathsf{S}}_{00}}{\tau^{\text{LP}}}, \label{195048_13Dec24}
 \end{align}
\label{eq: effective relaxation matrix first line}%
\end{subequations}
where $\tau^{\text{AM}}$ and $\tau^{\text{LP}}$ are introduced as relaxation times.
\begin{subequations}%
The relaxation rate of the CPs, i.e., phonon AM, becomes
 \begin{equation}
 \frac{1}{\tau^{\text{AM}}(q)} = \frac{2}{3}\left(c_{\text{L}}-c_{\text{T}}\right)^2q^2 \tau_2, \label{120637_5Sep24}
 \end{equation}
while the LPs follow another relaxation rate
 \begin{equation}
 \frac{1}{\tau^{\text{LP}}(q)} = \frac{2}{5}\left(c_{\text{L}}-c_{\text{T}}\right)^2q^2 \tau_2. \label{120358_2Dec24}
 \end{equation}
\label{142356_28Nov24}%
\end{subequations}
Remarkably, the relaxation times $\tau^{\text{AM}}$ and $\tau^{\text{LP}}$ are much longer than the impurity relaxation time $\tau_2$,
since $(c_{\text{L}} - c_{\text{T}}) q \tau_2 \ll 1$. 
In particular, LPs relax more slowly than CPs.
Furthermore, $\tau^{\text{AM}}$ and $\tau^{\text{LP}}$ are inversely proportional to $\tau_2$.
This confirms that phonon polarizations relax more slowly in crystals with greater impurity content.
We can also derive the relaxation rates~\eqref{120637_5Sep24}\eqref{120358_2Dec24}
by evaluating the near-zero eigenvalues of the matrix in Eq.~\eqref{112102_16Aug24} through second-order perturbation theory.

We can generalize the above result. 
Since the operation ${\Omega}\Omega^{\dagger}$ or $\widetilde{{\Omega}}\widetilde{{\Omega}}^\dagger$
in Eqs.~\eqref{195042_13Dec24}\eqref{195048_13Dec24} corresponds to double commutators with ${H}_{\bm{q}}$, 
the second-order perturbation theory reduces to a formula~\cite{Dyakonov1972,Fabian2007}
\begin{equation}
\frac{\partial \Braket{{\rho}_{\bm{q}}} }{\partial t} \simeq -
\frac{\tau^{*}}{\hbar^2}
\Braket{
\left[{H}_{\bm{q}}, \left[{H}_{\bm{q}}, \Braket{{\rho}_{\bm{q}}}
\right]
\right]},
\label{eq: double commutator DP formula}
\end{equation}
which is independent of Hamiltonian details.
Here, the bracket denotes the angular average over $\hat{\bm{q}}$. 
Instead of $\tau_2$, we approximated the formula by using a typical impurity relaxation time $\tau^*(\overline{\omega}_q)$ 
as $\tau^{*}\equiv \max\{\tau_l\}_{l = 1, 2, \dots}$~\footnote{
In this approximation, we also regard the linear-momentum relaxation time $\tau_1$ as $\tau^{*}$.
}. 
We note that Eq.~\eqref{eq: double commutator DP formula} holds in the overdamped regime,
where all elements of ${H}^{\prime}_{\bm{q}}\tau^* / \hbar$ are perturbatively small. 
When any element approaches or exceeds 1, the motional narrowing breaks down.
We then expect that polarization dynamics exhibits damped oscillations,
analogous to electron spin relaxation~\cite{Gridnev2001,Burkov2004,Zutic2004,Culcer2007,Winkler2008,Boross2013,Kiss2016,Szolnoki2017,Szolnoki2017a,Cosacchi2017}.

\textit{\textcolor[named]{DarkGreen}{Relaxation rates in a cubic crystal}}.---Our motional narrowing framework applies to crystals 
with isotropic defects~\cite{Klemens1955,Tamura1983,Srivastava1990,ZimanTextbookElPh}
where the rate $1/\tau^*$ of the polarization-conserving scattering (i) satisfies ${H}^{\prime}_{\bm{q}}\tau^* / \hbar\ll 1$ and 
(ii) is higher than that of polarization-flip scattering~\footnote{
A typical example of the polarization-flip scattering is a three-phonon process due to lattice anharmonicity~\cite{PeierlsTextbook,ZimanTextbookElPh,Srivastava1990}. 
}.
We now estimate the relaxation rates of polarizations in solid argon (${^{36}}$Ar) 
with isotropic defects introduced by mass substitutions in its fcc lattice. 
Solid argon serves as an ideal model system: it has only acoustic modes and electron scattering is negligible~\cite{Christen1975}.

Unlike isotropic elastic bodies, the five-fold degeneracy of the relaxation rates $1/\tau^{\text{LP}}$ [see Eq.~\eqref{195048_13Dec24}] 
will split into two values,
$1/\tau^{\text{LP-1}}$ for polarizations $a = 3, 8$ and $1/\tau^{\text{LP-2}}$ for $a = 1, 4, 6$. 
This splitting occurs because the ${m}\bar{3}{m}$ symmetry of the fcc lattice couples the polarizations 
$a = 3, 8$ and $a = 1, 4, 6$ within their respective groups, 
but not between different groups (see also Fig.~\ref{fig: phon pol table and motional narr}).

\begin{figure}[tbp]
 \centering
\includegraphics[pagebox=artbox,width=0.90\columnwidth]{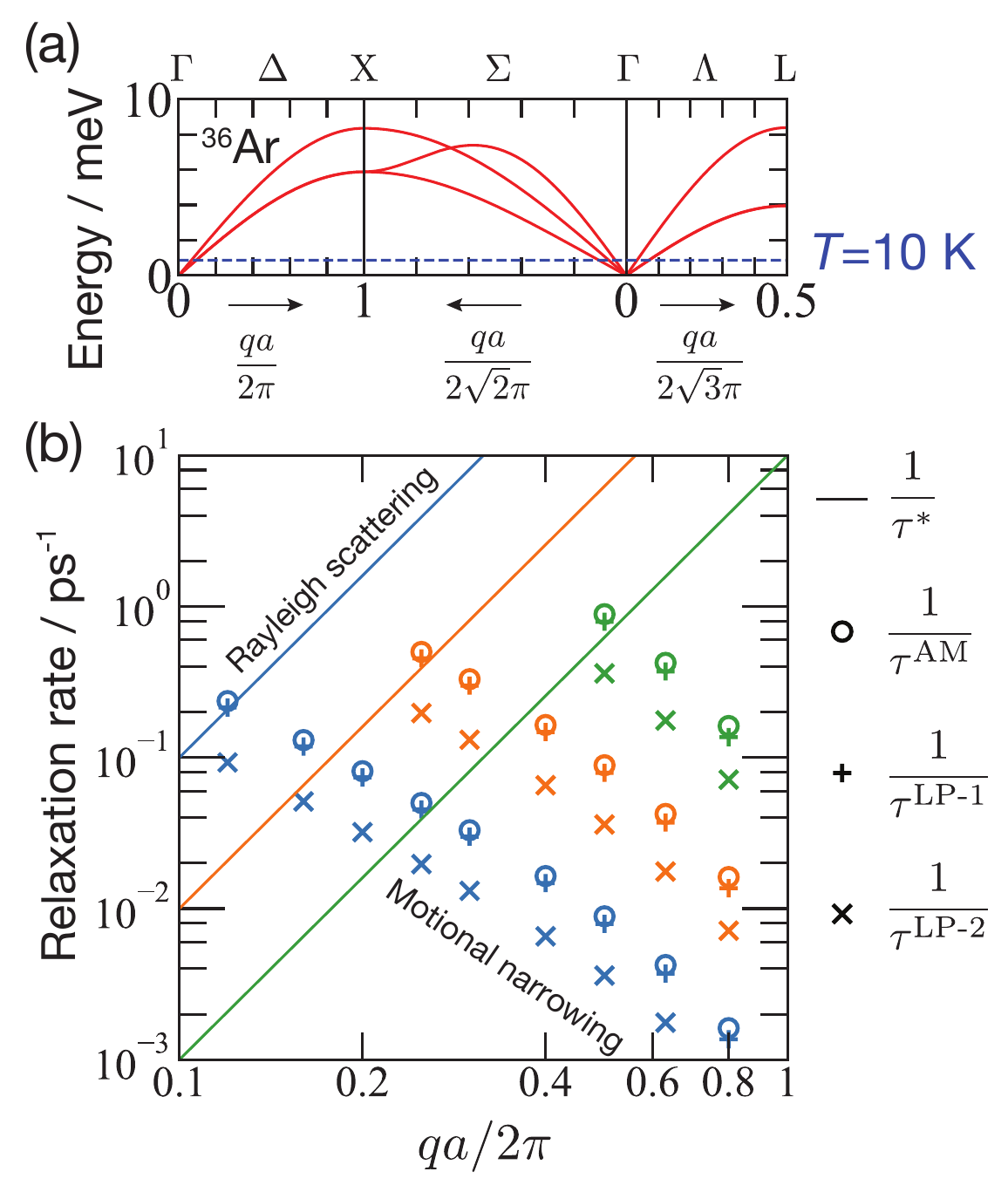}
\caption{
Phonon properties and relaxation dynamics in disordered solid argon.
(a)~Phonon dispersions at $\SI{10}{K}$. 
(b)~Relaxation rates $1/\tau^{\text{AM}}$, $1/\tau^{\text{LP-1}}$, and $1/\tau^{\text{LP-2}}$
of phonon polarizations as functions of wavevector $q$.
Three colors represent disorder strengths: 
$\alpha\cdot (2\pi/a)^4 =$ \SI{1e1}{ps^{-1}} (green), \SI{1e2}{ps^{-1}} (orange), and \SI{1e3}{ps^{-1}} (blue).
We also show the impurity relaxation rate $1/\tau^* = \alpha q^{4}$ due to the Rayleigh scattering by isotropic scatterers. 
}
\label{fig: solidAr disp and rel times vs qa}
\end{figure}

Figure~\ref{fig: solidAr disp and rel times vs qa}(a) shows the phonon dispersions at $\SI{10}{K}$, based on interatomic force constants 
from model~2 in Ref.~\cite[Table~II]{Fujii1974}.
With the formula~\eqref{eq: double commutator DP formula} and the Hamiltonian derived from the force constants,
we have calculated the relaxation rate of polarizations~\cite{SupplementalMaterial}.

The results are shown in Fig.~\ref{fig: solidAr disp and rel times vs qa}(b).  
The colored lines represent the impurity relaxation rate $1/\tau^*(q)$, 
which we modeled as $1/\tau^* = \alpha q^4$ on the basis of the Rayleigh scattering~\cite{ZimanTextbookElPh,Srivastava1990}. 
Then, the overdamped condition ${H}^{\prime}_{\bm{q}}\tau^* / \hbar \ll 1$ holds at large $q$. 
In this regime, motional narrowing yields three relaxation rates as expected: 
$1/\tau^{\text{LP-1}}$ for LPs with $a = 3, 8$; $1/\tau^{\text{LP-2}}$ for LPs with $a = 1, 4, 6$; and $1/\tau^{\text{AM}}$ for CPs. 
These rates, plotted in Fig.~\ref{fig: solidAr disp and rel times vs qa}(b), are noticeably lower than $1/\tau^*$ for large $q$. 
We also find that LPs relax more slowly than CPs.  

Figure~\ref{fig: solidAr disp and rel times vs qa}(b) displays the relaxation rates for three values of $\alpha$, distinguished by three colors. 
We observe that as the disorder strength $\alpha$ increases, the rates $1/\tau^{\text{LP-1}}$, $1/\tau^{\text{LP-2}}$, and $1/\tau^{\text{AM}}$ decrease. 
Thus, crystal disorder impedes the relaxation of polarizations.

We next examine the common $q$ dependence of $1/\tau^{\text{LP-1}}$, $1/\tau^{\text{LP-2}}$, and $1/\tau^{\text{AM}}$, 
collectively denoted by $1/\tau^{\text{Pol}}$. 
As shown in Eqs.~\eqref{120637_5Sep24}\eqref{120358_2Dec24}, $1/\tau^{\text{Pol}}$ is proportional to both the square of the phonon-band splitting $\hbar\Omega(q)$ 
and the impurity relaxation time $\tau^*(q)\,(\propto q^{-4})$, 
i.e., $1/\tau^{\text{Pol}} \propto \Omega^2(q) q^{-4}$.  
Since the band splitting $\hbar\Omega(q)$ scales linearly with $q$ at small $q$ and becomes constant at large $q$ [Fig.~\ref{fig: solidAr disp and rel times vs qa}(a)], 
$1/\tau^{\text{Pol}} \propto q^{-2}$ at small $q$ and $q^{-4}$ at large $q$. 
This $q$ dependence well explains that of three relaxation rates of polarizations in Fig.~\ref{fig: solidAr disp and rel times vs qa}(b).
We conclude from Fig.~\ref{fig: solidAr disp and rel times vs qa} that 
phonons with larger $q$ have remarkably longer lifetimes of polarizations and dominate the transport of AM and LP.
Conversely, long-wavelength phonons contribute to heat transport, 
since the scattering rate $1/\tau^*$ of linear momentum vanishes at small $q$.

Finally, we estimate the relaxation rates for a hypothetical solid mixture of ${}^{36}\text{Ar}_{1-x}{}^{136}\text{Xe}_x$ with a ratio of mass substitution $x = 6\%$. 
On the basis of a conventional model~\cite{Klemens1955,Tamura1983,Srivastava1990},
the disorder strength is calculated as $\alpha\cdot (2\pi/a)^4 =$ \SI{28}{ps^{-1}}~\cite{SupplementalMaterial},
which lies between the green and orange lines in Fig.~\ref{fig: solidAr disp and rel times vs qa}(b). 
We find that the relaxation times of polarizations exceed the impurity relaxation time at $qa/(2\pi) \simeq 0.4$, which corresponds to \SI{40}{K}. 
The phonon distribution, slightly deviated from equilibrium at \SI{10}{K}, usually has moderate weight around \SI{40}{K}.
These estimations thus indicate a potential possibility for observing long-lived phonon polarization in this system.

\textit{\textcolor[named]{DarkGreen}{Discussions and conclusions}}.---We have discovered a mechanism where
crystal disorder hinders the relaxation of phonon polarizations.
This finding contrasts with a na\"ive expectation that greater disorder would accelerate the relaxation.
We have identified the mechanism as motional narrowing:
polarization oscillations due to phonon-band splitting change to overdamped slow dynamics under frequent scattering. 
In disordered crystals, this mechanism enhances the relaxation time of phonon AM.
Furthermore, we have characterized LPs by five distinctive parameters and found that they relax more slowly than the AM via the same mechanism.

It is important to recognize that 
the AM addressed in this Letter---phonon spin AM~\cite{Vonsovskii1962,Levine1962,Ishii1975,Mclellan1988,Zhang2014}---is not conserved in crystals. 
Even in isotropic elastic bodies, i.e., crystals with isotropic elastic properties, the AM is not a constant but oscillates over time.
Moreover, we have confirmed that
the AM relaxes to zero over a long but finite timescale $\tau^{\text{AM}}$ under isotropic scatterers without directional or polarization dependencies.

Our study reveals the motional narrowing mechanism for all the $3\times 3$-polarization degrees of freedom in phonons.
This approach, focusing on all polarization components, will also apply to the dynamics of electron orbital degrees of freedom~\cite{Han2022,Sohn2024}. 
We expect that not only orbital AM of electrons~\cite{Sohn2024} but also 
its symmetrized products, orbital angular positions, exhibit the motional narrowing.

A practical method to observe the slow decay of phonon polarization is to measure diffusion length of locally generated AM or LP of phonons. 
The diffusion lengths are estimated as $l^{\text{AM}} = \sqrt{D\tau^{\text{AM}}}$ and $l^{\text{LP}} = \sqrt{D\tau^{\text{LP}}}$,
where $D = v^2\tau^*/3$ depends on the sound velocity $v$ and impurity relaxation time $\tau^*$.
Since $\tau^{\text{AM}}, \tau^{\text{LP}} \propto 1/\tau^{*}$, the diffusion lengths are independent of $\tau^*$.
This independence would be tested experimentally by varying sample thickness and impurity concentration.

\begin{acknowledgments}
{We wish to thank E.~Minamitani, A.~Shitade, K.~Yoshimi, S.~Sumita, and Y.~Kato for helpful discussions.
In particular, we thank E.~Minamitani for communication on calculating phonon relaxation times.
This work was supported by JSPS KAKENHI Grant Numbers, JP22H00108, JP22K18687,
JP24H02231 and JP24KJ1036, and also by MEXT Initiative to Establish Next-generation Novel Integrated Circuits Centers (X- NICS) Grant Number JPJ011438.}
\end{acknowledgments}
%

\clearpage
\renewcommand{\thesection}{S\arabic{section}}
\renewcommand{\theequation}{S\arabic{equation}}
\setcounter{equation}{0}
\renewcommand{\thefigure}{S\arabic{figure}}
\setcounter{figure}{0}
\renewcommand{\thetable}{S\arabic{table}}
\setcounter{table}{0}

\makeatletter
\newsavebox{\@brx}
\newcommand{\llangle}[1][]{\savebox{\@brx}{\(\m@th{#1\langle}\)}%
  \mathopen{\copy\@brx\mkern2mu\kern-0.9\wd\@brx\usebox{\@brx}}}
\newcommand{\rrangle}[1][]{\savebox{\@brx}{\(\m@th{#1\rangle}\)}%
  \mathclose{\copy\@brx\mkern2mu\kern-0.9\wd\@brx\usebox{\@brx}}}
\makeatother

\makeatletter
\c@secnumdepth = 2
\makeatother
\onecolumngrid
\begin{center}
 {\large \textmd{Supplemental Materials for:} \\[0.3em] {\bfseries Disorder-Induced Slow Relaxation of Phonon Polarization}} \\[1.0em]
{Yuta Suzuki and Shuichi Murakami}
\end{center}
\setcounter{page}{1}
\setcounter{section}{0}

\section{Gell-Mann matrices}
The eight Gell-Mann matrices are given as:  
\begin{subequations}
 \begin{align}
 {\lambda}_1 &= 
 \begin{bmatrix}
 0 & 1& 0\\ 1 & 0 & 0 \\ 0 & 0 & 0
 \end{bmatrix} ,
 &{\lambda}_2 &= 
 \begin{bmatrix}
 0 & -i & 0 \\ i & 0 & 0 \\ 0 & 0 & 0 
 \end{bmatrix},
 &{\lambda}_3 &= 
 \begin{bmatrix}
 1 & 0 & 0 \\ 0 & -1 & 0 \\ 0 & 0& 0
 \end{bmatrix},\\
 {\lambda}_4 &=
 \begin{bmatrix}
 0 & 0 & 1 \\ 0 & 0 & 0 \\ 1 & 0 & 0
 \end{bmatrix},
 &{\lambda}_5 &=
 \begin{bmatrix}
 0 & 0 & -i \\ 0 & 0 & 0 \\ i & 0 & 0
 \end{bmatrix}, & & \\
 {\lambda}_6 &=
 \begin{bmatrix}
 0 & 0& 0 \\ 0 & 0& 1 \\ 0 & 1 & 0
 \end{bmatrix},
 & {\lambda}_7 &=
 \begin{bmatrix}
 0 & 0& 0 \\ 0 & 0& -i \\ 0 & i & 0
 \end{bmatrix},
 &{\lambda}_8 &= \frac{1}{\sqrt{3}}
 \begin{bmatrix}
 1 & 0 & 0 \\ 0 & 1 & 0 \\ 0 & 0 & -2
 \end{bmatrix}.
 \end{align}
\end{subequations}
These matrices are traceless, $\tr[{\lambda}_a] = 0$, and satisfy the orthonormality condition $\tr[{\lambda}_a{\lambda}_b] = 2\delta_{ab}$.  
The commutators between these matrices can be expressed as:
\begin{equation}
 \left[{\lambda}_a, {\lambda}_b\right] = 2i \sum_{c = 1}^8 f_{abc} {\lambda}_c,
\end{equation}
where the structure constants $f_{abc}$ are antisymmetric with respect to the indices $a$, $b$, and $c$.
The independent nonzero elements of $f_{abc}$ are: 
\begin{align}
 f_{123}&=1, &
 f_{147}&=f_{246}=f_{257}=f_{345}=\frac{1}{2}, &
 f_{156}&=f_{367}=-\frac{1}{2}, &
 f_{458}&= f_{678}= \frac{\sqrt{3}}{2}. \label{113550_5Dec24}
\end{align}
It is worth noting that $f_{abc}$ vanishes if
all three indices $a$, $b$, and $c$ belong to $\mathsf{S}\equiv \{1, 3, 4, 6, 8\}$ or 
if two indices are in $\mathsf{AS}\equiv \{2, 5, 7\}$ and one in $\mathsf{S}$.
We have used this property to derive Eqs.~(2a)(2b) in the main text.

\section{Symmetry properties of phonon Hamiltonian}
In Eq.~(1) of the main text, the phonon Hamiltonian ${H}_{\bm{q}}$ is symmetric, and therefore
expressed as a linear combination of the Gell-Mann matrices ${\lambda}_a$
with $a \in \mathsf{S} \equiv \left\{1, 3, 4, 6, 8\right\}$ and the identity matrix. The reason is as follows:

The equation of motion for the plane-wave displacement $2\mathrm{Re}[\bm{u} \exp (i \bm{q}\cdot \bm{r} - i\omega t)]$, 
discussed in the main text, results in an eigenvalue problem
${D}(\bm{q}) \bm{u}  = \omega^2\bm{u}$, where ${D}(\bm{q})$ is the dynamical matrix.
Here, we assume that the crystal is nonmagnetic and centrosymmetric.
Then, the dynamical matrix ${D}(\bm{q})$ satisfies three conditions: 
\begin{enumerate}
 \item ${D}(\bm{q})$ is a positive semi-definite Hermitian matrix, ${D}^{\dagger}(\bm{q}) = {D}(\bm{q})$.
 \item Time-reversal symmetry requires a reciprocity ${D}(\bm{q}) = {D}^{\top}(-\bm{q})$.
 \item For centrosymmetric crystals, ${D}(-\bm{q}) = {D}(\bm{q})$.
\end{enumerate}

These conditions ensure that $D = {D}(\bm{q})$ is a real symmetric matrix.
We can thus diagonalize it as $D = O \Delta O^{\top}$.
Here, $O$ is a real orthogonal matrix, and $\Delta$ is a real diagonal matrix.  
Since $D$ is positive semi-definite, all diagonal elements of $\Delta$ are nonnegative:  
\begin{equation}  
\Delta =  
\begin{bmatrix}  
\delta_1 & & \\  
& \delta_2 & \\  
& & \delta_3  
\end{bmatrix},  
\qquad \delta_i \geq 0 \quad \text{for} \quad i = 1, 2, 3.  
\end{equation}  
We define $\Delta^{1/2}$ by:  
\begin{equation}  
\Delta^{1/2} =  
\begin{bmatrix}  
\sqrt{\delta_1} & & \\  
& \sqrt{\delta_2} & \\  
& & \sqrt{\delta_3}  
\end{bmatrix}.  
\end{equation}  
Note that $\Delta^{1/2}$ is a real positive semi-definite matrix. 
The Hamiltonian $H = H_{\bm{q}}$ is then given by $H = \hbar O \Delta^{1/2} O^{\top}$.
This matrix $H$ is symmetric ($H^{\top} = H$), real, and has nonnegative eigenvalues.  
In particular, we can expand $H_{\bm{q}}$
using symmetric matrices ${\lambda}_a$ with $a \in \mathsf{S}$ and $I_3$.

\section{Invariants in undamped dynamics of polarization}

We have illustrated oscillation of polarization in the absence of scattering in Eqs.~(2a)(2b) of the main text.
Here, we analyze this motion by identifying conserved quantities.

From the von~Neumann equation $\partial {\rho}_{\bm{q}}/\partial t = - ({i}/{\hbar})\left[{H}^{\prime}_{\bm{q}},{\rho}_{\bm{q}}\right]$,
we find that the trace is conserved:
\begin{equation}
\frac{\partial}{\partial t}\tr \left[\rho_{\bm{q}}\right]
= -\frac{i}{\hbar} \tr\left[\left[H^{\prime}_{\bm{q}} ,\rho_{\bm{q}} \right]\right] = 0.
\end{equation}
Similarly, multiplying both sides by $\rho_{\bm{q}}$ and taking the trace shows that: 
\begin{equation}
\frac{\partial}{\partial t}\left(\frac{1}{2}\tr \left[\rho_{\bm{q}}^2\right]\right)
= \tr \left[\rho_{\bm{q}}\frac{\partial \rho_{\bm{q}}}{\partial t} \right]
= -\frac{i}{\hbar}
\tr\left[ \rho_{\bm{q}} \left[H^{\prime}_{\bm{q}} ,\rho_{\bm{q}} \right]\right] = 0.
\end{equation}
Thus, $\tr \left[\rho_{\bm{q}}\right]$ and $\tr \left[\rho_{\bm{q}}^2\right]/2$ are conserved.

Meanwhile, we can expand the $3\times 3$ density matrix $\rho_{\bm{q}}$ as:
\begin{equation}
 \rho_{\bm{q}} = \frac{1}{3}\left(\Lambda_0\cdot I_3 + \sum_{a=1}^8\Lambda_a\lambda_a\right),
\end{equation}
where $\Lambda_0 \equiv \tr[\rho_{\bm{q}}]$ and $\Lambda_a \equiv (3/2) \tr[\rho_{\bm{q}}\lambda_a]$.
We then find that $\sum_{a = 1}^8 {\Lambda_{a}}^2$ is invariant. This follows from an equality
\begin{equation}
 \frac{1}{2}\tr \left[\rho^2_{\bm{q}}\right] = 
\frac{1}{18}
\tr\left[
{\Lambda_0}^2\cdot I_3 + \Lambda_0\sum_{a=1}^8\Lambda_a\lambda_a
+ \sum_{a=1}^8\sum_{b=1}^8\Lambda_a\Lambda_b\cdot \lambda_a\lambda_b\right]
= \frac{1}{6} \left(\tr \left[\rho_{\bm{q}}\right]\right)^2 + \frac{1}{9}\sum_{a = 1}^8 {\Lambda_{a}}^2.
\end{equation}
We conclude that the vector $(\Lambda_1,\Lambda_2, \dots, \Lambda_8)$ moves on a hypersphere of radius 
$\sqrt{\sum_{a = 1}^8 \Lambda^2_a}$ embedded in the eight-dimensional Euclidean space.

\section{Definition of phonon angular momentum}

In footnote~[56] of the main text, we noted that phonon angular momentum~(AM) off equilibrium includes an interband term
other than the expectation value of the spin-1 operator. Here, we provide a detailed expression and discuss its relaxation.

We consider low-energy acoustic phonons in a crystal with $N$ unit cells, 
(averaged) atomic mass $M$, and atomic lattice points $\bm{R}_{j}$ with displacements $\hat{\bm{u}}_{j}$. 
The displacement $\hat{\bm{u}}_{j}$ is expanded as:
\begin{equation}
 \hat{\bm{u}}_{j}(t)
= \sum_{\bm{q}, n}e^{i( \bm{q}\cdot \bm{R}_{j}-\omega_{\bm{q}n} t)} 
\sqrt{\frac{\hbar}{2NM\omega_{\bm{q}n}}} \hat{a}_{\bm{q}n}\cdot \bm{e}_{\bm{q}n} + \text{H.c.},
\label{154042_3Sep24}
\end{equation}
where $\hbar\omega_{\bm{q}n}$, $\bm{e}_{\bm{q}n}$, and $\hat{a}_{\bm{q}n}$ represent
the phonon energy, polarization vector, and annihilation operator for a mode $(\bm{q}, n)$, respectively. 

We define the phonon (spin) AM as the statistical average of the mechanical AM of all atoms~\cite{Vonsovskii1962,Levine1962,Ishii1975,Mclellan1988,Zhang2014}:
\begin{align}
& \bm{\ell}^{\text{mec}} = \sum_{j =1}^N {\llangle[\Bigg] \hat{\bm{u}}_{j}\times M
\frac{d \hat{\bm{u}}_{j}}{d t}\rrangle[\Bigg]} \nonumber \\
&= {- i \hbar}\sum_{\bm{q}, n, n'}
(\bm{e}^{*}_{\bm{q}n'}\times \bm{e}_{\bm{q}n}) e^{- i  (\omega_{\bm{q}n}- \omega_{\bm{q}n'})t}
\left[
\frac{1}{2}\left(\sqrt{\frac{\omega_{\bm{q}n'}}{\omega_{\bm{q}n}}} + \sqrt{\frac{\omega_{\bm{q}n}}{\omega_{\bm{q}n'}}}\right)
{\llangle\hat{a}^{\dagger}_{\bm{q}n'}\hat{a}_{\bm{q}n}\rrangle} 
+ \frac{\delta_{nn'}}{2}
\right].\label{163708_23Aug24}
\end{align}
Here, we expressed the statistical average as the double angle bracket.

In time-reversal symmetric crystals, 
the last term proportional to $\delta_{nn'}$ vanishes because the dynamical matrix satisfies $D^{*}(\bm{q}) = D(-\bm{q})$, allowing us to set 
$\bm{e}^{*}_{\bm{q}n} = \bm{e}_{-\bm{q},n}$ under an appropriate $\mathrm{U}(1)$ gauge. 
As a result, the factor 
$\sum_{n'}(\bm{e}^{*}_{\bm{q}n'}\times \bm{e}_{\bm{q}n}) e^{- i  (\omega_{\bm{q}n}- \omega_{\bm{q}n'})t} \delta_{nn'}
= \bm{e}_{-\bm{q},n} \times \bm{e}_{\bm{q},n}$ is odd in $\bm{q}$. 
The summation of this odd term over the Brillouin zone is zero. 

The remaining terms of $\bm{\ell}^{\text{mec}}$ can be expressed using the phonon density matrix:
\begin{equation}
 \left(\varrho_{\bm{q}} \right)_{\mu, \nu} (t)\equiv e^{- i  (\omega_{\bm{q}n}- \omega_{\bm{q}n'})t}
\sum_{n, n'} e_{\bm{q}n, \mu}{\llangle\hat{a}^{\dagger}_{\bm{q}n'}\hat{a}_{\bm{q}n}\rrangle} e^{*}_{\bm{q}n',\nu}.
\end{equation}
Let us decompose $\bm{\ell}^{\text{mec}}$ into two terms:
\begin{subequations} %
 \begin{gather}
 \bm{\ell}^{\text{mec}} = \bm{\ell}^{(0)} + \bm{\ell}^{\text{corr}},\\
 \begin{aligned}
 \bm{\ell}^{(0)}&= \sum_{\bm{q}} \tr \left[ \bm{S} \varrho_{\bm{q}}\right], \label{113818_3Sep24}\\
 \bm{\ell}^{\text{corr}}& = {- i \hbar}\sum_{\bm{q}, n, n'}
 (\bm{e}^{*}_{\bm{q}n'}\times \bm{e}_{\bm{q}n}) 
 \frac{1}{2}\left(\sqrt{\frac{\omega_{\bm{q}n'}}{\omega_{\bm{q}n}}} + \sqrt{\frac{\omega_{\bm{q}n}}{\omega_{\bm{q}n'}}} -2\right)
 {\llangle\hat{a}^{\dagger}_{\bm{q}n'}\hat{a}_{\bm{q}n}\rrangle} e^{- i  (\omega_{\bm{q}n}- \omega_{\bm{q}n'})t}\\
 &= \sum_{\bm{q}}
 \sum_{\substack{n, n'\\ (\omega_{\bm{q}n}\neq \omega_{\bm{q}n'})}}
 \sum_{\substack{\mu, \nu, \kappa, \sigma\\ = x, y, z }}
 {e}^{*}_{\bm{q}n',\mu} \left(\bm{S}\right)_{\mu, \nu} {e}_{\bm{q}n,\nu}
 \cdot \frac{1}{2}\left(\sqrt{\frac{\omega_{\bm{q} n'}}{\omega_{\bm{q} n}}} + \sqrt{\frac{\omega_{\bm{q} n}}{\omega_{\bm{q} n'}}}-2\right)
 \cdot e^{*}_{\bm{q}n,\kappa} \left(\varrho_{\bm{q}} \right)_{\kappa,\sigma} e_{\bm{q}n',\sigma},
 \end{aligned}
 \end{gather}
\end{subequations}
where ${\bm{S}} = ({S}_x, {S}_y, {S}_z)$ represents spin-1 operators defined as $({S}_{\mu})_{\nu, \kappa}= -i\hbar \epsilon_{\mu\nu\kappa}$ ($\mu, \nu, \kappa = x, y, z$).
While $\bm{\ell}^{(0)}$ measures circular polarization~(CP), 
$\bm{\ell}^{\text{corr}}$ is an interband correction arising off equilibrium.

For isotropic elastic bodies characterized by acoustic modes $\omega_{\bm{q}\text{L}} = c_{\text{L}}q$ and $\omega_{\bm{q}\text{T}} = c_{\text{T}}q$
with sound velocities $c_{\text{L}}$ (longitudinal) and $c_{\text{T}}$ (transverse), the interband correction reduces to:
\begin{align}
& \bm{\ell}^{\text{corr}} = \frac{1}{2}\left(\sqrt{\frac{c_{\text{T}}}{c_{\text{L}}}} + \sqrt{\frac{c_{\text{L}}}{c_{\text{T}}}} -2 \right)
\sum_{\bm{q}} \sum_{\mu, \nu, \kappa, \sigma}  
\left(\bm{S}\right)_{\mu,\nu}
\left(\varrho_{\bm{q}} \right)_{\kappa,\sigma} \nonumber\\
& \qquad \times \left[
\Big(
\sum_{\substack{n' \\ (\omega_{\bm{q}n'} = \omega_{\bm{q}\text{T}})}} e_{\bm{q}n',\sigma}{e}^{*}_{\bm{q}n',\mu}
\Big)\cdot {e}_{\bm{q}\text{L},\nu} e^{*}_{\bm{q}\text{L},\kappa}
 + e_{\bm{q}\text{L},\sigma} {e}^{*}_{\bm{q}\text{L},\mu} \cdot
\Big(\sum_{\substack{n \\ (\omega_{\bm{q}n} = \omega_{\bm{q}\text{T}})}} 
{e}_{\bm{q}n,\nu} e^{*}_{\bm{q}n,\kappa} \Big) \right]\nonumber\\
& = \frac{1}{2}\left(\sqrt{\frac{c_{\text{T}}}{c_{\text{L}}}} + \sqrt{\frac{c_{\text{L}}}{c_{\text{T}}}} -2 \right)
\sum_{\bm{q}} \sum_{\mu, \nu, \kappa, \sigma}  \left(\bm{S}\right)_{\mu,\nu}
\left(\varrho_{\bm{q}} \right)_{\kappa,\sigma} \nonumber\\
& \qquad \times \left[
\left(
\delta_{\sigma\mu} - e_{\bm{q}\text{L},\sigma}{e}^{*}_{\bm{q}\text{L},\mu}
\right)\cdot {e}_{\bm{q}\text{L},\nu} e^{*}_{\bm{q}\text{L},\kappa}
 + e_{\bm{q}\text{L},\sigma} {e}^{*}_{\bm{q}\text{L},\mu} 
\left(
\delta_{\nu\kappa} - {e}_{\bm{q}\text{L},\nu} e^{*}_{\bm{q}\text{L},\kappa} 
\right)\right].\label{112605_3Sep24}
\end{align}
By using the polarization of the longitudinal mode $\bm{e}_{\bm{q}\text{L}} = i\hat{\bm{q}}$, we obtain
\begin{align}
{\ell}^{\text{corr}}_{\mu}
= \sum_{\bm{q}} \sum_{\kappa, \sigma} 
\frac{-i\hbar}{2}\left(\sqrt{\frac{c_{\text{T}}}{c_{\text{L}}}} + \sqrt{\frac{c_{\text{L}}}{c_{\text{T}}}} -2 \right)\cdot
\left(
\sum_{\nu} \epsilon_{\mu\sigma\nu} \hat{q}_{\nu}\hat{q}_{\kappa}
+ \sum_{\chi} \epsilon_{\mu\chi\kappa}\hat{q}_{\sigma} \hat{q}_{\chi}
\right)\cdot
\left(\varrho_{\bm{q}} \right)_{\kappa,\sigma}.
\end{align}

Thus, the total mechanical AM can be expressed compactly as:
\begin{equation}
 \bm{\ell}^{\text{mec}} = \bm{\ell}^{(0)} + \bm{\ell}^{\text{corr}} = 
\sum_{\bm{q}} \tr [\bm{\mathcal{S}}\varrho_{\bm{q}}],
\label{181335_4Dec24}
\end{equation}
where $\bm{\mathcal{S}} = (\mathcal{S}_x,\mathcal{S}_y,\mathcal{S}_z)$ incorporates both the spin-1 operator and corrections:
\begin{align}
&\left(\mathcal{S}_{\mu}\right)_{\sigma,\kappa}
 = \left({S}_{\mu}\right)_{\sigma,\kappa}
- \frac{i\hbar}{2}\left(\sqrt{\frac{c_{\text{T}}}{c_{\text{L}}}} + \sqrt{\frac{c_{\text{L}}}{c_{\text{T}}}} -2 \right)\cdot
\left(
\sum_{\nu} \epsilon_{\mu\sigma\nu} \hat{q}_{\nu}\hat{q}_{\kappa}
+ \sum_{\chi} \epsilon_{\mu\chi\kappa}\hat{q}_{\sigma} \hat{q}_{\chi}
\right)\nonumber\\
 &= \frac{1}{3}\left(\sqrt{\frac{c_{\text{T}}}{c_{\text{L}}}} + \sqrt{\frac{c_{\text{L}}}{c_{\text{T}}}} + 1 \right) \left({S}_{\mu}\right)_{\sigma,\kappa}
- \frac{i\hbar}{2}
\left(\sqrt{\frac{c_{\text{T}}}{c_{\text{L}}}} + \sqrt{\frac{c_{\text{L}}}{c_{\text{T}}}} -2 \right)
\sum_{\chi, \nu, \tau} \epsilon_{\mu\chi\nu}
\left(
\delta_{\nu\kappa}\delta_{\sigma\tau} - \delta_{\nu\sigma}\delta_{\kappa\tau}
\right)
\left(
\hat{q}_{\chi}\hat{q}_{\tau} - \frac{\delta_{\chi\tau}}{3} 
\right).\label{120043_5Sep24}
\end{align}
Since $\bm{\ell}^{\text{mec}} = 0$ in equilibrium, we can replace $\varrho_{\bm{q}}$ with $\rho_{\bm{q}} = \varrho_{\bm{q}} - \varrho^{\text{eq}}_{\bm{q}}$
in Eq.~\eqref{181335_4Dec24}.

Finally, $\bm{\ell}^{\text{mec}}$ relaxes to zero with two rates. 
The first term of Eq.~\eqref{120043_5Sep24}, proportional to $\left({S}_{\mu}\right)_{\sigma,\kappa}$,
relaxes slowly with a rate $1/\tau^{\text{AM}}$, while the second term involving second-order spherical harmonics of ${\rho}_{\bm{q}}$ relaxes faster with a rate
$1/\tau_2~(\gg 1/\tau^{\text{AM}})$. 
Thus, the long-term relaxation is governed by the slower rate $1/\tau^{\text{AM}}$, 
regardless of whether $\bm{\ell}^{\text{corr}}$ is included in the definition of AM.

\section{Expansion of quantum kinetic equation}

In Eqs.~(5) and (7) of the main text, we expand the quantum kinetic equation:  
\begin{equation}
 \frac{\partial {\rho}_{\bm{q}}}{\partial t} + \frac{i}{\hbar}\left[{H}^{\prime}_{\bm{q}} ,{\rho}_{\bm{q}}\right] 
= \frac{2\pi}{\hbar} \sum_{\bm{q}'}\mathcal{K}_{\bm{q},\bm{q}'}
\left({\rho}_{\bm{q}'} - {\rho}_{\bm{q}}\right) \cdot \delta (\hbar \overline{\omega}_{q} -\hbar \overline{\omega}_{q'}).
\label{092712_5Dec24}
\end{equation}
for isotropic elastic bodies
with respect to both the spherical harmonics $Y_{lm}(\hat{\bm{q}})$ and the Gell-Mann matrices ${\lambda}_a$.
In Eq.~\eqref{092712_5Dec24}, the commutator term involves ${H}^{\prime}_{\bm{q}}$ shown in Eq.~(4) of the main text, i.e, 
\begin{equation}
\left({H}^{\prime}_{\bm{q}}\right)_{\mu\nu}
= \hbar (c_{\text{L}} - c_{\text{T}}) q\left(\hat{q}_{\mu}\hat{q}_{\nu} - \frac{\delta_{\mu\nu}}{3}\right),
 \label{093328_5Dec24}
\end{equation}
while $\overline{\omega}_{q} = vq$.  
In this section, we follow the calculation of the expansion, and 
derive explicit expressions for the expansion coefficients $\Omega^{a;b}_{lm;l'm'}$, 
time constants $\tau_l$, and the submatrices ${\Omega}$, $\widetilde{{\Omega}}$, and ${{\Omega}'}$, introduced in the main text.  

We start by expanding the density matrix as:
\begin{equation}
 {\rho}_{\bm{q}}(t) = {\rho}^0_{\bm{q}}(t)\cdot I_3
+ \sum_{a = 1}^8\sum_{l=0}^{\infty}\sum_{m=-l}^l {\rho}^{a}_{lm}(q, t) Y_{lm}(\hat{\bm{q}}){\lambda}_a \label{103413_6Dec24}
\end{equation}
with coefficients ${\rho}^{a}_{lm}(q, t)$.
We then expand the time-derivative term in Eq.~\eqref{092712_5Dec24} in the same way.
Similarly, we will expand the commutator on the left-hand side and the collision integral on the right-hand side of Eq.~\eqref{092712_5Dec24}.

\subsection{Expansion of the commutator}

We expand the traceless part $H^{\prime}_{\bm{q}}$, as defined in Eq.~\eqref{093328_5Dec24}, 
using second-order spherical harmonics $Y_{2m}(\hat{\bm{q}})$
and symmetric matrices $\lambda_a$ with $a\in \mathsf{S}\equiv \{1, 3, 4, 6, 8\}$:
\begin{equation}
\hat{q}_{\mu}\hat{q}_{\nu} - \frac{\delta_{\mu\nu}}{3}
= \sum_{a \in \mathsf{S}}\sum_{m=-2}^2 \gamma^{a}_{2m} Y_{2m}(\hat{\bm{q}}) \left({\lambda}_{a} \right)_{\mu\nu}.
\label{231226_8Jul24}
\end{equation}
The expansion coefficients $\gamma^{a}_{2m}$ are uniquely determined by the orthonormality of spherical harmonics and Gell-Mann matrices. The nonzero components are:  
\begin{equation}
\gamma^{a}_{2m} =
\begin{cases}
\displaystyle
-i\sqrt{\frac{2\pi}{15}} & (a, m) = (1, +2), \\[10pt] 
\displaystyle
+i\sqrt{\frac{2\pi}{15}} & (a, m) = (1, -2), (6,-1), (6, +1)\\[10pt] 
\displaystyle
+\sqrt{\frac{2\pi}{15}} & (a, m) = (3, -2), (3, +2), (4, -1)\\[10pt] 
\displaystyle
-\sqrt{\frac{2\pi}{15}} & (a, m) = (4, +1)\\[10pt] 
\displaystyle
-\sqrt{\frac{4\pi}{15}} & (a, m) = (8, 0)\\[10pt] 
\end{cases}.
\label{113537_5Dec24}
\end{equation}
All other components vanish. From Eq.~\eqref{113537_5Dec24}, we identify a symmetry relation:  
\begin{equation}
 \gamma^a_{2, -m} = (-1)^m \left(\gamma^a_{2m}\right)^{*}.\label{101100_6Dec24}
\end{equation}

Using $\gamma^{a}_{2m}$, we expand the commutator in the quantum kinetic equation~\eqref{092712_5Dec24} as follows:
\begin{align}
&\frac{i}{\hbar}\left[H^{\prime}_{\bm{q}} ,\rho_{\bm{q}}\right]
= i (c_{\text{L}}-c_{\text{T}}) q \sum_{a_1, m_1} \sum_{a_2, l_2, m_2}
\gamma^{a_1}_{2 m_1} \rho^{a_2}_{l_2 m_2}(q)
Y_{2 m_1}(\hat{\bm{q}}) Y_{l_2m_2}(\hat{\bm{q}})\left[{\lambda}_{a_1},{\lambda}_{a_2}\right]\nonumber\\
&= -2 (c_{\text{L}}-c_{\text{T}}) q \sum_{a, l,  m}
\sum_{a_1, m_1} \sum_{a_2, l_2, m_2}
\gamma^{a_1}_{2 m_1} 
f_{a a_1 a_2} 
(-1)^{m} \sqrt{\frac{5(2l+1)(2l_2+1)}{4\pi}}\nonumber\\
&\qquad \times
\begin{pmatrix}
 l & 2 & l_2 \\ 0 & 0 & 0
\end{pmatrix}
\begin{pmatrix}
 l & 2 & l_2 \\ -m & m_1 & m_2
\end{pmatrix}
\rho^{a_2}_{l_2 m_2}(q) Y_{l,m}(\hat{\bm{q}}){\lambda}_a \nonumber\\
&\equiv - \sum_{a, l, m} Y_{lm}(\hat{\bm{q}}) \lambda_a \sum_{b, l', m'} i\Omega^{a;b}_{lm;l'm'} {\rho}^{b}_{l'm'},
\label{110957_6Dec24}
\end{align}
where we reduce the product of spherical harmonics using the Wigner 3j symbol:
\begin{equation}
 \int  d  \hat{\bm{q}}~Y^{*}_{lm}(\hat{\bm{q}})Y_{l_1 m_1}(\hat{\bm{q}})Y_{l_2m_2}(\hat{\bm{q}})
= (-1)^{m}\sqrt{\frac{(2l+1)(2l_1 +1)(2l_2 +1)}{4\pi}}
\begin{pmatrix}
 l & l_1 & l_2 \\ 0 & 0 & 0
\end{pmatrix} 
\begin{pmatrix}
 l & l_1 & l_2\\ -m & m_1 & m_2
\end{pmatrix}.
\label{100737_6Dec24}
\end{equation}
Here, $\int  d  \hat{\bm{q}}$ denotes angular integration over the full solid angle.
The coefficient $\Omega^{a;b}_{lm;l'm'}$, defined by Eq.~\eqref{110957_6Dec24}, is given by:
\begin{equation}
 i\Omega^{a;b}_{lm;l'm'} \equiv 2(c_{\text{L}}-c_{\text{T}}) q 
\sum_{c \in \mathsf{S}} \sum_{\widetilde{m} = -2}^2
\gamma^{c}_{2\widetilde{m}} f_{a c b} (-1)^{m} \sqrt{\frac{5(2l+1)(2l'+1)}{4\pi}}
\begin{pmatrix}
 l & 2 & l' \\ 0 & 0 & 0
\end{pmatrix}
\begin{pmatrix}
 l & 2 & l' \\ -m & \widetilde{m} & m'
\end{pmatrix}. \label{101133_6Dec24}
\end{equation}
The coefficient $\Omega^{a;b}_{lm;l'm'}$ vanishes if any of the following conditions hold:
\begin{itemize}
 \item $a, b\in \mathsf{S}$,
 \item $a, b\in \mathsf{AS}$,
 \item $\Abs{l-l'} > 2$,
 \item $l + l' <2$, or
 \item $l + l'$ is odd.
\end{itemize}
This is because:
\begin{enumerate}
 \item The structure constant $f_{acb}$ with $c \in \mathsf{S}$ becomes zero for $a, b \in \mathsf{S}$ or $a, b \in \mathsf{AS}$
[see also Eq.~\eqref{113550_5Dec24}], and
 \item The Wigner 3j symbol $\begin{pmatrix}  l & 2 & l' \\ 0 & 0 & 0 \end{pmatrix}$ vanishes 
unless triangle inequality $\Abs{l-l'}\leq 2 \leq l+l'$ holds and $l + l'$ is even.
\end{enumerate}
Moreover, the coefficient $\Omega^{a;b}_{lm;l'm'}$ satisfies a symmetry relation:
\begin{equation}
 \Omega^{b;a}_{l'm';lm} = \left(\Omega^{a;b}_{lm;l'm'} \right)^{*},\label{101622_6Dec24}
\end{equation}
which is derived as follows:
\begin{align}
 & i\Omega^{b;a}_{l'm';lm}=2(c_{\text{L}}-c_{\text{T}}) q \sum_{c, \widetilde{m}} 
\gamma^{c}_{2\widetilde{m}} f_{b c a} (-1)^{m'} \sqrt{\frac{5(2l+1)(2l'+1)}{4\pi}}
\begin{pmatrix}
 l' & 2 & l \\ 0 & 0 & 0
\end{pmatrix}
\begin{pmatrix}
 l' & 2 & l \\ -m' & \widetilde{m} & m
\end{pmatrix}\nonumber\\
&= 2(c_{\text{L}}-c_{\text{T}}) q \sum_{c, \widetilde{m}} 
\gamma^{c}_{2, -\widetilde{m}}\cdot (-f_{a c b})\cdot  (-1)^{m'} \sqrt{\frac{5(2l+1)(2l'+1)}{4\pi}}
\begin{pmatrix}
 l' & 2 & l \\ 0 & 0 & 0
\end{pmatrix}
\begin{pmatrix}
 l' & 2 & l \\ -m' & -\widetilde{m} & m
\end{pmatrix}\nonumber\\
&= -2(c_{\text{L}}-c_{\text{T}}) q \sum_{c, \widetilde{m}} 
\left(\gamma^{c}_{2, \widetilde{m}}\right)^* f_{a c b} (-1)^{m' + \widetilde{m}} \sqrt{\frac{5(2l+1)(2l'+1)}{4\pi}}
\begin{pmatrix}
 l' & 2 & l \\ 0 & 0 & 0
\end{pmatrix}
\begin{pmatrix}
 l' & 2 & l \\ m' & \widetilde{m} & -m
\end{pmatrix}\nonumber\\
 &= -2(c_{\text{L}}-c_{\text{T}}) q \sum_{c, \widetilde{m}} 
\left(\gamma^{c}_{2, \widetilde{m}}\right)^* f_{a c b} (-1)^{m} \sqrt{\frac{5(2l+1)(2l'+1)}{4\pi}}
\begin{pmatrix}
 l & 2 & l' \\ 0 & 0 & 0
\end{pmatrix}
\begin{pmatrix}
 l & 2 & l' \\ -m & \widetilde{m} & m'
\end{pmatrix}\nonumber\\
 &= -\left(i\Omega^{a;b}_{lm;l'm'}\right)^{*}.
\end{align}
Here, we used Eqs.~\eqref{101100_6Dec24} and \eqref{101133_6Dec24}, $l+l'\in 2 \mathbb{Z}$,
and the properties of the Wigner 3j symbols:
\begin{subequations}
 \begin{gather}
 \begin{pmatrix}
 l_1 & l_2 & l_3 \\ -m_1 & -m_2 & -m_3
 \end{pmatrix}
 = (-1)^{l_1 + l_2 + l_3}
 \begin{pmatrix}
 l_1 & l_2 & l_3 \\ m_1 & m_2 & m_3
 \end{pmatrix}
 = 
 \begin{pmatrix}
 l_3 & l_2 & l_1 \\ m_3 & m_2 & m_1
 \end{pmatrix}, \\
 \begin{pmatrix}
 l_1 & l_2 & l_3 \\ m_1 & m_2 & m_3
 \end{pmatrix} = 0 \qquad\quad \text{when}\quad
 m_1 + m_2 + m_3 \neq 0.
\end{gather}
\end{subequations}

\subsection{Expansion of the collision integral}

The scattering rate $\mathcal{K}_{\bm{q},\bm{q}'}$ depends only on the modulus $q = q'$ and the relative angle between $\hat{\bm{q}}$ and $\hat{\bm{q}}'$.
We can expand it using the Legendre polynomials:
\begin{equation}
\mathcal{K}_{\bm{q},\bm{q}'}
= \sum_{l=0}^{\infty} (2l +1) \mathcal{K}_l(q) \cdot P_l(\cos\vartheta)
= \sum_{l=0}^{\infty} \sum_{m=-l}^l
4\pi \mathcal{K}_l(q) Y^{*}_{lm}(\hat{\bm{q}'}) Y_{lm}(\hat{\bm{q}}),
\end{equation}
where $\vartheta$ is the angle between $\hat{\bm{q}}$ and $\hat{\bm{q}}'$, $P_l(\cos\vartheta)$ is the Legendre polynomial of order $l$,
and $\mathcal{K}_l(q)$ is the expansion coefficient.

Substituting the expansion of $\rho_{\bm{q}}$ shown in Eq.~\eqref{103413_6Dec24} into the collision integral yields:
\begin{align}
 &\frac{2\pi}{\hbar} \sum_{\bm{q}'}\mathcal{K}_{\bm{q},\bm{q}'}
\left(\rho_{\bm{q}'} - \rho_{\bm{q}}\right) \cdot \delta (\hbar v q -\hbar vq')\nonumber\\
&= \frac{2\pi}{\hbar} g(\overline{\omega}_q)
\left\{
\int \frac{d\hat{\bm{q}'}}{4\pi} \mathcal{K}_{\bm{q}, \bm{q}'}
\left[\rho^0_{\bm{q}'}\cdot I_3
+ \sum_{a = 1}^8\sum_{l=0}^{\infty}\sum_{m=-l}^l \rho^{a}_{lm}(q) Y_{lm}(\hat{\bm{q}'}){\lambda}_a
\right]\right.\nonumber\\
&\qquad  -  \left.\left.\left[\rho^0_{\bm{q}}\cdot I_3
+ \sum_{a = 1}^8\sum_{l=0}^{\infty}\sum_{m=-l}^l \rho^{a}_{lm}(q) Y_{lm}(\hat{\bm{q}}){\lambda}_a\right] 
\int \frac{d\hat{\bm{q}'}}{4\pi} \mathcal{K}_{\bm{q}, \bm{q}'}
\right\}\right|_{q' = q}\nonumber\\
& = \frac{2\pi}{\hbar} g(\overline{\omega}_q) \left[
\left.\int \frac{ d\hat{\bm{q}}}{4\pi} \mathcal{K}_{\bm{q},\bm{q}'}\left(\rho^0_{\bm{q}'}-\rho^0_{\bm{q}}\right)\right|_{q' =q} \cdot I_3
+ \sum_{a,l,m} \mathcal{K}_l(q)  \rho^{a}_{lm}(q) Y_{lm}(\hat{\bm{q}}){\lambda}_a
- \sum_{a,l,m} \mathcal{K}_0(q)  \rho^{a}_{lm}(q) Y_{lm}(\hat{\bm{q}}){\lambda}_a \right]\nonumber\\
&\equiv \frac{2\pi}{\hbar} g(\overline{\omega}_q) 
\left.\int \frac{ d\hat{\bm{q}}}{4\pi} \mathcal{K}_{\bm{q},\bm{q}'}\left(\rho^0_{\bm{q}'}-\rho^0_{\bm{q}}\right)\right|_{q' =q} \cdot I_3
- \sum_{a = 1}^8\sum_{l=0}^{\infty}\sum_{m= -l}^l 
\frac{1}{\tau_l (\overline{\omega}_q)} \rho^{a}_{lm}(q) Y_{lm}(\hat{\bm{q}}){\lambda}_a.
\label{111013_6Dec24}
\end{align}
Here, $g(\overline{\omega}_q = vq) = \sum_{\bm{q}'} \delta (\hbar v q -\hbar vq') = V {\overline{\omega}_q}^2/(2\pi^2\hbar v^3)$
denotes the density of states with $V$ the volume of the crystal.
The relaxation time $\tau_l$ for the $l$-th spherical harmonics of $\rho_{\bm{q}}$ is defined as:
\begin{equation}
\frac{1}{\tau_0 (\overline{\omega}_q)} \equiv 0,\qquad 
\frac{1}{\tau_l (\overline{\omega}_q)}\equiv \frac{2\pi}{\hbar} g(\overline{\omega}_q)
\int_{-1}^1 \frac{d (\cos\vartheta)}{2}
\mathcal{K}_{\bm{q},\bm{q}'}\left[1 - P_l(\cos\vartheta)\right] > 0 \quad (l = 1,2,3,\dots).
\end{equation}

\subsection{Simplification of the quantum kinetic equation}

We expand the quantum kinetic equation~\eqref{092712_5Dec24} in the basis $\{Y_{lm}(\hat{\bm{q}})\lambda_a\}$ 
using Eqs.~\eqref{103413_6Dec24}, \eqref{110957_6Dec24}, and \eqref{111013_6Dec24}.
Then, we obtain the equation of motion for $\rho^a_{lm}$ as follows:
\begin{equation}
\frac{\partial {\rho}^{a}_{lm}}{\partial t}
= \sum_{b, l', m'} i\Omega^{a;b}_{lm;l'm'} {\rho}^{b}_{l'm'}  - \frac{{\rho}^a_{lm}}{\tau_l},
\quad \text{where}\quad
\begin{cases}
a = 1, 2,\dots, 8, \\
l= 0,1,2\dots,\\
m= -l, -l+1,\dots, l,
\end{cases}
\label{114050_6Dec24}
\end{equation}
which is the same as Eq.~(5) in the main text.

Since this equation is linear in ${\rho}^{a}_{lm}$, we can arrange it into matrix form.
As noted in the paragraph following Eq.~\eqref{101133_6Dec24}, 
the coefficient $\Omega^{a;b}_{lm;l'm'}$ is subject to some constraints.
Based on the constraints, we rewrite a part of Eq.~\eqref{114050_6Dec24} as follows:
\begin{equation}
 \frac{\partial}{\partial t} 
 \left[
\renewcommand\arraystretch{1.3}
\begin{array}{c}
 \bm{\rho}^{\mathsf{AS}}_{00} \\[10pt]
 \bm{\rho}^{\mathsf{AS}}_{2} \\[10pt]
 \bm{\rho}^{\mathsf{AS}}_{\mathrm{extra}} \\[10pt] \hline
 \bm{\rho}^{\mathsf{S}}_{00} \\[10pt]
 \bm{\rho}^{\mathsf{S}}_{2} \\[10pt]
 \bm{\rho}^{\mathsf{S}}_{\mathrm{extra}} 
\end{array} 
\right]  = 
 \left[
\renewcommand\arraystretch{1.3}
\begin{array}{ccc|ccc}
  O                 & O                    & O                          & O                               & i\Omega                   & O                            \\[10pt]
  O                 & -\dfrac{1}{\tau_2}   & O                          & - i\widetilde{\Omega}^{\dagger}  & i\Omega'                  & i\Omega''                     \\[10pt]
  O                 & O                    & -\dfrac{1}{\tau_{l= 1,3,\dots}}& O                               & i {\Omega'''}^{\dagger}  & i\Omega''''                   \\[10pt] \hline
  O                 & i\widetilde{\Omega}   & O                          & O                               & O                        & O                            \\[10pt]
  i\Omega^{\dagger} & i{\Omega'}^{\dagger}& i\Omega'''                  & O                               & -\dfrac{1}{\tau_2}       & O                            \\[10pt]
  O                 & i{\Omega''}^{\dagger}& i{\Omega''''}^{\dagger}   & O                               & O                        & -\dfrac{1}{\tau_{l= 1,3,\dots}}
\end{array} 
\right] \left[
\renewcommand\arraystretch{1.3}
\begin{array}{c}
 \bm{\rho}^{\mathsf{AS}}_{00} \\[10pt]
 \bm{\rho}^{\mathsf{AS}}_{2} \\[10pt]
 \bm{\rho}^{\mathsf{AS}}_{\mathrm{extra}} \\[10pt] \hline
 \bm{\rho}^{\mathsf{S}}_{00} \\[10pt]
 \bm{\rho}^{\mathsf{S}}_{2} \\[10pt]
 \bm{\rho}^{\mathsf{S}}_{\mathrm{extra}} 
\end{array} 
\right],
 \end{equation}
which is a generalized form of Eq.~(7) in the main text.
Here, the submatrix $\dfrac{1}{\tau_{l= 1,3,\dots}}$ denotes a diagonal matrix with elements $1/\tau_1$, $1/\tau_3$, $1/\tau_4$, \dots.
We introduced vectors $\bm{\rho}^{\mathsf{AS}}_{00}(q, t) = (\rho^2_{00}, \rho^5_{00}, \rho^7_{00})^{\top}$,
$\bm{\rho}^{\mathsf{S}}_{00}(q, t) = (\rho^1_{00}, \rho^3_{00}, \rho^4_{00}, \rho^6_{00}, \rho^8_{00})^{\top}$, and 
\begin{align}
\bm{\rho}^{\mathsf{S}}_{2}&
=  \left[
\renewcommand\arraystretch{1.3}
\begin{array}{c}
\bm{R}_1 \\\hline
\bm{R}_3 \\\hline
\bm{R}_4 \\\hline
\bm{R}_6 \\\hline
\bm{R}_8
\end{array} 
\right],
&\bm{\rho}^{\mathsf{AS}}_{2}
&=  \left[
\renewcommand\arraystretch{1.3}
\begin{array}{c}
\bm{R}_2 \\\hline
\bm{R}_5 \\\hline
\bm{R}_7
\end{array} 
\right] 
\end{align}
with
\begin{subequations} 
 \begin{gather}
 \begin{aligned}
 \bm{R}_1&=
 \begin{bmatrix}
 \rho^{1}_{2, -2} \\[8pt]
 \rho^{1}_{2, -1} \\[8pt]
 \rho^{1}_{2, +1} \\[8pt]
 \rho^{1}_{2, +2} 
 \end{bmatrix},
 &\bm{R}_3&=
 \begin{bmatrix}
 \rho^{3}_{2, -2} \\[8pt]
 \rho^{3}_{2, -1} \\[8pt]
 \rho^{3}_{2, +1} \\[8pt]
 \rho^{3}_{2, +2}  
 \end{bmatrix},
 &\bm{R}_4&=
 \begin{bmatrix}
 \rho^{4}_{2, -2} \\[8pt]
 \rho^{4}_{2, -1} \\[8pt]
 \rho^{4}_{2, 0} \\[8pt]
 \rho^{4}_{2, +1} \\[8pt]
 \rho^{4}_{2, +2}  
 \end{bmatrix},
 &\bm{R}_6&=
 \begin{bmatrix}
 \rho^{6}_{2, -2} \\[8pt]
 \rho^{6}_{2, -1} \\[8pt]
 \rho^{6}_{2, 0} \\[8pt]
 \rho^{6}_{2, +1} \\[8pt]
 \rho^{6}_{2, +2}
 \end{bmatrix},
 &\bm{R}_8&=
 \begin{bmatrix}
 \rho^{8}_{2, -1} \\[8pt]
 \rho^{8}_{2, +1} 
 \end{bmatrix}
 \end{aligned},\\
 \begin{aligned}
 \bm{R}_2&=
 \begin{bmatrix}
 \rho^{2}_{2, -2} \\[8pt]
 \rho^{2}_{2, -1} \\[8pt]
 \rho^{2}_{2, +1} \\[8pt]
 \rho^{2}_{2, +2} 
 \end{bmatrix},
 &\bm{R}_5&=
 \begin{bmatrix}
 \rho^{5}_{2, -2} \\[8pt]
 \rho^{5}_{2, -1} \\[8pt]
 \rho^{5}_{2, 0} \\[8pt]
 \rho^{5}_{2, +1} \\[8pt]
 \rho^{5}_{2, +2}  
 \end{bmatrix},
 &\bm{R}_7&=
 \begin{bmatrix}
 \rho^{7}_{2, -2} \\[8pt]
 \rho^{7}_{2, -1} \\[8pt]
 \rho^{7}_{2, 0} \\[8pt]
 \rho^{7}_{2, +1} \\[8pt]
 \rho^{7}_{2, +2}  
 \end{bmatrix}
 \end{aligned}.
\end{gather}
\end{subequations}
The other vectors $\bm{\rho}^{\mathsf{AS}}_{\mathrm{extra}}$ and $\bm{\rho}^{\mathsf{S}}_{\mathrm{extra}}$
consist of components $\rho^a_{lm}$ with $l = 1,3,4\dots$.
We discarded these components in Eq.~(7) of the main text, since
these components do not couple directly to the dynamics of $\rho^a_{00}$
whereas they may couple to $\bm{\rho}^{\mathsf{AS}/\mathsf{S}}_{2}$.

We now provide the elements of the key submatrices $\Omega$ and $\widetilde{\Omega}$, according to Eq.~\eqref{101133_6Dec24}:
\begin{align}
i\Omega
&= \frac{\left(c_{\text{L}}-c_{\text{T}}\right)q}{\sqrt{30}}
 \left[
\renewcommand\arraystretch{1.3}
\begin{array}{c|c|c|c|c}
M_1 & M_3 & M_4 & M_6 & M_8
\end{array} 
\right],
& i\widetilde{\Omega}
&= \frac{\left(c_{\text{L}}-c_{\text{T}}\right)q}{\sqrt{30}}
 \left[
\renewcommand\arraystretch{1.3}
\begin{array}{c|c|c}
M_2 & M_5 & M_7
\end{array} 
\right],
\end{align}
where
\begin{subequations}
 \begin{align}
 M_1& =  
 \begin{bmatrix}
 2 & 0   & 0   & 2 \\[8pt]
 0 & i & i & 0 \\[8pt]
 0 & -1  & 1   & 0 
 \end{bmatrix},
 &M_3 &= 
 \begin{bmatrix}  
 2i & 0    & 0	   & -2i \\[8pt]
 0    & -1   & 1	   & 0     \\[8pt]
 0    & -i & -i & 0     
 \end{bmatrix},
 &M_4 &= 
 \begin{bmatrix}
 0	  & i & 0	      & i & 0   \\[8pt]
 1	  & 0	  & -\sqrt{6} & 0   & 1   \\[8pt]
 -i & 0	  & 0	      & 0   & i 
 \end{bmatrix},\\
 M_6& =  
 \begin{bmatrix}
 0     & 1 & 0	 & -1 & 0  \\[8pt]
 -i  & 0 & 0	 & 0  & i\\[8pt]
 -1    & 0 & -\sqrt{6} & 0  & -1
 \end{bmatrix},
 &M_8 &= 
 \begin{bmatrix}
 0		  & 0  \\[8pt]
 -\sqrt{3}	  & \sqrt{3} \\[8pt]
 \sqrt{3}i & \sqrt{3}i 
 \end{bmatrix},
 \\
 M_2& =  
 \begin{bmatrix}
 -2 & 0   & 0   & -2 \\[8pt]
 -2i & 0& 0 & 2i \\[8pt]
 0 & -i  & -i   & 0 \\[8pt]
 0 & -1  & 1   & 0 \\[8pt]
 0 & 0  & 0   & 0 
 \end{bmatrix},
 &M_5 &= 
 \begin{bmatrix}  
 0 & -i   & 0	   & -i&0 \\[8pt]
 0    & 1   & 0	   & -1   & 0 \\[8pt]
 -1    &  0& \sqrt{6} & 0     & -1\\[8pt]
 i & 0  & 0   & 0 & -i  \\[8pt]
 0 & \sqrt{3}  & 0   & -\sqrt{3} & 0
 \end{bmatrix},
 &M_7 &= 
 \begin{bmatrix}
 0 & 1& 0 & -1 & 0   \\[8pt]
 0 & i	 & 0 & i  & 0   \\[8pt]
 i & 0& 0 & 0   & -i \\[8pt]
 1 & 0  & \sqrt{6}   & 0 & 1\\[8pt]
 0 & -\sqrt{3}i  & 0   & -\sqrt{3}i & 0
 \end{bmatrix}.
 \end{align}
\end{subequations}
Thus, we have the following relation shown in Eqs.~(8a)--(9b) of the main text:
\begin{subequations} 
 \begin{align}
 \Omega \Omega^{\dagger} &= \frac{\left(c_{\text{L}}-c_{\text{T}}\right)^2q^2}{30}
 \sum_{a \in \mathsf{S}} M_a M_a^{\dagger}
 =\frac{2}{3}\left(c_{\text{L}}-c_{\text{T}}\right)^2q^2, \\
 \widetilde{\Omega} \widetilde{\Omega}^{\dagger} & = \frac{\left(c_{\text{L}}-c_{\text{T}}\right)^2q^2}{30}
 \sum_{a \in \mathsf{AS}} M_a M_a^{\dagger}
 = \frac{2}{5}\left(c_{\text{L}}-c_{\text{T}}\right)^2q^2.
 \end{align}
\end{subequations}

\section{Calculation of polarization relaxation rates in solid argon}

In this section, we present a method to calculate the relaxation rates of the statistical average of phonon polarization from the formula~(10) in the main text.
Figure~3(b) in the main text displays the computed relaxation rates $1/\tau^{\text{AM}}$, $1/\tau^{\text{LP-1}}$, and $1/\tau^{\text{LP-2}}$ for solid argon 
using this method.

We focus on the isotropic part $\Braket{{\rho}_{\bm{q}}}$ of the density matrix ${{\rho}_{\bm{q}}}$, 
where the bracket denotes an angular average over the wavevector direction $\hat{\bm{q}}$. 
The dynamics of $\Braket{{\rho}_{\bm{q}}}$ is governed by:
\begin{equation}
\frac{\partial \Braket{{\rho}_{\bm{q}}} }{\partial t} \simeq -
\frac{\tau^{*}}{\hbar^2}
\Braket{
\left[{H}_{\bm{q}}, \left[{H}_{\bm{q}}, \Braket{{\rho}_{\bm{q}}}
\right]
\right]}\label{235151_19Dec24}
\end{equation}
at a fixed modulus of the wavevector $q$,
which is the same as Eq.~(10) in the main text.

With the help of Eq.~\eqref{103413_6Dec24}, we decompose $\Braket{{\rho}_{\bm{q}}}$ into components $\rho^a_{00}(q, t)$ ($a = 1, 2,\dots, 8$)
that are independent of $\hat{\bm{q}}$:
\begin{equation}
 \Braket{\rho_{\bm{q}}(t)}
= \Braket{\rho^0_{\bm{q}}}\cdot I_3 + \frac{1}{\sqrt{4\pi}} \sum_{a =1}^8 \rho^a_{00}(q, t) \lambda_a.
\end{equation}

Substituting this decomposition and the expansion of the Hamiltonian
\begin{equation}
{H}_{\bm{q}} = h_0(\bm{q})\cdot{I}_3 + \sum_{a \in \mathsf{S}} h_a (\bm{q}) {\lambda}_a, 
\quad\text{where}\quad 
h_a (\bm{q}) = \frac{1}{2}\tr \left[H_{\bm{q}} \lambda_a\right],
\end{equation}
into Eq.~\eqref{235151_19Dec24}, we derive the equation of motion for each $\rho^a_{00}(q, t)$:
\begin{align}
\frac{\partial \rho^a_{00} (q, t)}{\partial t}  &= - \frac{\tau^*}{2\hbar}
\sum_{b =1}^8\sum_{c, d\in\mathsf{S}}
\Braket{\tr\Big[
\left[h_c(\bm{q}) \lambda_c, 
\left[h_d(\bm{q}) \lambda_d, \rho^b_{00}(q, t) \lambda_b\right]\right]\lambda_a\Big]}\nonumber\\
 &= 
\begin{cases}
\displaystyle  - \frac{\tau^*}{\hbar}\sum_{b\in\mathsf{S}}\sum_{c, d\in\mathsf{S}}\sum_{e\in\mathsf{AS}}
\Braket{h_c (\bm{q}) h_d (\bm{q})} f_{ace} f_{edb} \rho^b_{00}(q, t) & a\in \mathsf{S},\\[4pt]
\displaystyle - \frac{\tau^*}{\hbar}\sum_{b\in\mathsf{AS}}\sum_{c, d\in\mathsf{S}}\sum_{e\in\mathsf{S}}
\Braket{h_c (\bm{q}) h_d (\bm{q})} f_{ace} f_{edb} \rho^b_{00}(q, t) & a\in \mathsf{AS}
\end{cases}
\label{002858_20Dec24}\\ 
&\equiv
\begin{cases}
\displaystyle - \sum_{b\in \mathsf{S}} \left[\Gamma^{\mathsf{S}}(q)\right]_{ab} \rho^b_{00}(q, t) & a\in \mathsf{S},\\[4pt]
\displaystyle - \sum_{b\in \mathsf{AS}} \left[\Gamma^{\mathsf{AS}}(q)\right]_{ab} \rho^b_{00}(q, t) & a\in \mathsf{AS}
\end{cases}.
\label{233925_19Dec24}
\end{align}
Here, $\Gamma^{\mathsf{S}}(q)$ and $\Gamma^{\mathsf{AS}}(q)$ are $5\times 5$ and $3\times 3$ matrices 
that describe the decay of $\rho^a_{00}(q, t)$. 
For the case of the solid argon, we numerically performed the angular integration over all solid angle, denoted by the brackets in Eq.~\eqref{002858_20Dec24}.

To simplify Eq.~\eqref{233925_19Dec24}, we introduce two vectors $\bm{\rho}^{\mathsf{AS}}_{00}(q, t) = (\rho^2_{00}, \rho^5_{00}, \rho^7_{00})^{\top}$ and 
$\bm{\rho}^{\mathsf{S}}_{00}(q, t) = (\rho^1_{00}, \rho^3_{00}, \rho^4_{00}, \rho^6_{00}, \rho^8_{00})^{\top}$.
This transforms Eq.~\eqref{233925_19Dec24} into:
\begin{equation}
 \frac{\partial}{\partial t} 
 \left[
\renewcommand\arraystretch{1.3}
\begin{array}{c}
 \bm{\rho}^{\mathsf{S}}_{00} \\ \hline
 \bm{\rho}^{\mathsf{AS}}_{00}
\end{array} 
\right]  \simeq -
 \left[
\renewcommand\arraystretch{1.3}
\begin{array}{c|c}
  \Gamma^{\mathsf{S}}                 & O      \\ \hline
  O                 & \Gamma^{\mathsf{AS}} 
\end{array} 
\right] \left[
\renewcommand\arraystretch{1.3}
\begin{array}{c}
 \bm{\rho}^{\mathsf{S}}_{00} \\ \hline
 \bm{\rho}^{\mathsf{AS}}_{00}
\end{array} 
\right].\label{000116_20Dec24}
\end{equation}
The relaxation dynamics of the polarization now reduces to an eigenvalue problem
of the $8 \times 8$ matrix on the right-hand side of Eq.~\eqref{000116_20Dec24}. 
Its eigenvalues provide the relaxation rates of $\rho^a_{00}(q, t)$, which 
correspond to the relaxation rates of the statistical average of phonon polarization 
at a fixed value of $q$ [see Eq.~(6) in the main text].
On the other hand, the eigenvectors specify the corresponding polarization states that decay at those rates.

\section{Estimation of mass-difference scattering rate}
We estimate the impurity relaxation rate $1/\tau^*$, caused by 
mass-difference scattering in a hypothetical solid mixture of ${}^{36}\text{Ar}_{1-x}{}^{136}\text{Xe}_x$ with $x = 6\%$. 
Based on the conventional model~\cite{Klemens1955,Tamura1983,Srivastava1990}, the scattering rate is expressed as:
\begin{equation}
 \frac{1}{\tau_{\text{md}}^* (\overline{\omega}_q)}
= \frac{\Gamma_{\text{md}}}{4\pi}\cdot \frac{V_{\mathrm{UC}}}{\overline{v}^3}\left( \overline{\omega}_q\right)^4,
\end{equation}
where $\overline{\omega}_q = \overline{v}q$ represents the averaged phonon dispersion,
$\overline{v}\simeq \SI{1e3}{m/s}$ is the averaged sound velocity, and 
$V_{\mathrm{UC}}$ is the volume of unit cell.
The $\omega^4$ dependence of $1/\tau_{\text{md}}^*$ reflects the Rayleigh scattering.

We calculate $V_{\mathrm{UC}}$ using the lattice constant $a = \SI{5.3}{\angstrom}$ of ${}^{36}\mathrm{Ar}$ at \SI{10}{K}, 
yielding $V_{\mathrm{UC}} = a^3/4$ for the fcc lattice.
The impurity scattering strength $\Gamma_{\text{md}}$ is defined as:
\begin{equation}
 \Gamma_{\text{md}}
= (1-x) \left(1- \frac{M_{\text{Ar}}}{\overline{M}} \right)^2
+ x \left(1- \frac{M_{\text{Xe}}}{\overline{M}} \right)^2,
\end{equation}
where $M_{\mathrm{Ar}}$ and $M_{\mathrm{Xe}}$ are the atomic masses of ${}^{36}\mathrm{Ar}$ and ${}^{136}\mathrm{Xe}$, respectively, 
and $\overline{M} = (1-x)M_{\mathrm{Ar}} + xM_{\mathrm{Xe}}$ is the mean atomic mass. For $x = 6\%$, we obtain $\Gamma_{\mathrm{md}} = 0.32$.

The relaxation rate can then be approximated as:
\begin{equation}
 \frac{1}{\tau_{\text{md}}^*(q)} \simeq \SI{2.8 }{ns^{-1}}\cdot\left(\frac{\hbar\overline{\omega}_q/k_{\text{B}}}{\SI{10}{K}}\right)^{4}
= \SI{2.8 }{ns^{-1}}\cdot\left(\frac{qa/2\pi}{0.1}\right)^4.
\label{142645_16Dec24}
\end{equation}
In other words, the expression for $1/\tau_{\text{md}}^*$ is given by $1/\tau_{\text{md}}^* = \alpha\cdot q^4$, where 
$\alpha \cdot \left(2\pi/a\right)^4 = \SI{28}{ns^{-1}}$.

\end{document}